\documentclass[onecolumn,pdflatex,sn-mathphys-num]{jicn}
\usepackage{graphicx,amsmath,amssymb,amsthm,mathtools}
\usepackage{booktabs,array,tabularx,multirow}
\usepackage[ruled,vlined,linesnumbered]{algorithm2e}
\usepackage{setspace,colortbl}
\usepackage{multicol,caption,subcaption}
\usepackage{times,mathptmx,microtype}
\usepackage{titlesec,geometry,fancyhdr}
\usepackage[english]{babel}
\usepackage{enumitem}
\usepackage{xurl}
\titleformat{\section}{\normalfont\fontsize{16}{17}\bfseries\raggedright\hyphenpenalty=10000}{\thesection}{.6em}{}
\titleformat{\subsection}{\normalfont\fontsize{12}{13}\bfseries\raggedright\hyphenpenalty=10000}{\thesubsection}{.6em}{}
\titleformat{\subsubsection}{\normalfont\fontsize{10.5}{12}\bfseries}{\thesubsubsection}{.6em}{}
\titlespacing{\section}{0pt}{.9em}{.65em}
\titlespacing{\subsection}{0pt}{.75em}{.5em}
\titlespacing{\subsubsection}{0pt}{.6em}{.35em}
\renewcommand{\thetable}{\Roman{table}}
\newenvironment{columntable}{\par\addvspace{8pt}\noindent\begin{minipage}{\columnwidth}\captionsetup{type=table}}{\end{minipage}\par\addvspace{9pt}}
\setlist{nosep,leftmargin=1.25em}
\hypersetup{colorlinks=true,urlcolor=blue,citecolor=blue,linkcolor=blue,pdftitle={Predictive Scheduling and Model Sharing for Continuous UAV Edge Inference},pdfauthor={Houyi Qi; Minghui Liwang}}
\theoremstyle{thmstyleone}
\newtheorem{proposition}{Proposition}
\newtheorem{lemma}{Lemma}
\newcommand{\ind}{\mathbf{1}}
\newcommand{\A}{\mathcal{A}}
\newcommand{\Sset}{\mathcal{S}}
\newcommand{\eps}{\varepsilon}
\newcommand{\SWAP}{SWAP}

\newcommand{\clip}{\operatorname{clip}}
\begin{document}
\thispagestyle{empty}

\begin{center}
	{\fontsize{20}{24}\selectfont\bfseries Predictive Scheduling and Model Sharing for Continuous UAV Edge Inference\par}
	\vspace{7pt}
	{\fontsize{11}{14}\selectfont Houyi Qi, Minghui Liwang$^{*}$\par}
	\vspace{4pt}
	{\fontsize{9}{11}\selectfont\itshape Shanghai Research Institute for Intelligent Autonomous Systems, Tongji University, Shanghai, China\par}
	\vspace{3pt}
	{\fontsize{9}{11}\selectfont
		E-mail: \href{mailto:houyiqi@tongji.edu.cn}{houyiqi@tongji.edu.cn} (Houyi Qi); \href{mailto:minghuiliwang@tongji.edu.cn}{minghuiliwang@tongji.edu.cn} (Minghui Liwang)\\
		$^{*}$Corresponding author: Minghui Liwang\par}
\end{center}

\noindent\textbf{Abstract:}
Continuous edge inference for unmanned aerial vehicles (UAVs) faces uncertain wireless connectivity, competition for shared computing resources, and the overhead of model loading and session-state migration. These interdependent factors complicate request admission and the continuity of successive inference services. This paper investigates joint request admission, edge-server assignment, and model selection under an admission-priority objective that strictly favors any feasible increase in admitted requests over bounded losses in delay, interruption risk, model quality, migration, and deployment. To address this problem, we propose Service-Window-Guided Admission and Model Placement (\SWAP), which integrates predictive service-window information with shared model deployment and coordinated scheduling. Quantile regression and probability calibration estimate the remaining service window and interruption risk, while a bounded residence factor characterizes the potential reuse of shared model instances. We then formulate a constrained exact potential game based on marginal-contribution utilities and develop a solution procedure combining unilateral improvement, coordinated admission, and finite binary logit refinement. Theoretical analysis establishes NP-hardness, the existence of a constrained pure Nash equilibrium, and finite descent, together with feasibility and approximate unilateral equilibrium guarantees for the returned assignment. Simulation results demonstrate improved service continuity and reduced request rejection under increasing resource contention, while characterizing the accompanying tradeoffs in model-deployment traffic, inference quality, and execution delay.

\vspace{5pt}
\vspace{8pt}

\begin{multicols}{2}
	\raggedcolumns

\section{Introduction}\label{sec:intro}

Unmanned aerial vehicles (UAVs) are increasingly employed in mobile perception, environmental monitoring, and other applications that require timely and continuous inference services. However, the limited onboard computing and storage resources of UAVs often constrain their ability to execute computationally intensive inference tasks. Ground-edge servers provide an effective alternative by hosting reusable inference models and sharing computing capacity among UAVs. Unlike isolated inference tasks, continuous UAV edge inference involves successive service requests under dynamically changing wireless connectivity and edge-resource availability. As UAVs move across different wireless coverage regions, a connection that offers sufficient transmission capacity at the decision instant may not persist until service completion. Meanwhile, switching the serving edge requires session-state migration, while selecting a nonresident inference model incurs cold-start delay and additional storage demand. These interdependent factors introduce significant challenges in maintaining service continuity and coordinating request admission, computing resources, and model availability.

Existing research has investigated various aspects of edge intelligence and resource coordination. Studies on UAV computing-power networks have explored joint model deployment and resource orchestration~\cite{cpn}, while heterogeneous edge inference has motivated research on model selection and configuration~\cite{huangdnn,qos}. Broader advances in distributed edge intelligence have also addressed communication-efficient foundation-model adaptation across heterogeneous AIoT devices~\cite{hou2026fedaf}. Furthermore, forecasting-driven service preparation and potential-game scheduling have been investigated for resource coordination in distributed air--ground edge networks~\cite{qi2026fusion}. To accommodate user mobility, existing approaches employ mobility-aware inference and service migration to adapt service placement~\cite{moei,rerouting}, whereas collaborative caching and computation offloading coordinate model availability with execution demands~\cite{collabcaching,twist}. Game-theoretic methods further address the interactions among offloading decisions under shared resource constraints~\cite{wcog,sotnra,pgevc,sgto}. Related studies have also explored hybrid futures--spot trading through cross-layer matching~\cite{qi2024bridge} and overbooking-based resource scheduling with reputation-guided contract updates~\cite{qi2025ohtrust}. Collectively, these studies provide important foundations for coordinating computing resources, model deployment, and mobile service provisioning in dynamic edge environments.

Despite these advances, continuous UAV edge inference requires closer coordination among request admission, wireless service continuity, and shared model deployment. Admitting an additional request may increase the computation delay of existing requests or require loading a new model instance, whereas a shared instance can serve multiple requests and provide reuse benefits across successive slots. Consequently, scheduling decisions based on individual service costs may overlook opportunities to accommodate additional requests through coordinated server assignments and model selections. Moreover, satisfying the current transmission-rate, model-quality, and execution-delay constraints does not necessarily ensure service completion, since wireless connections may be interrupted during execution. Addressing these challenges requires a joint scheduling framework that accounts for uncertain wireless service windows and the collective effects of computing contention and shared model deployment, while prioritizing the admission of feasible requests. Motivated by these considerations, we investigate the following research questions (RQs).

\noindent\textit{RQ 1: How can uncertain wireless service windows guide request admission and shared model deployment?}
The remaining connection duration is unknown at each decision instant, while shared model instances may serve multiple requests with different service-window characteristics. The key challenge is to exploit causal mobility and wireless observations to characterize interruption risk and potential model reuse, thereby supporting continuity-aware scheduling under physical service constraints.

\noindent\textit{RQ 2: How can request admission be prioritized under coupled computing and model-storage constraints?}
Admitting an additional request may increase the computation delay of existing requests or require loading a new model instance. Conversely, coordinated server assignments and model selections may release resources and accommodate additional requests. These interactions call for a joint optimization objective that captures shared resource costs while giving strict priority to any feasible increase in admitted request cardinality.

\noindent\textit{RQ 3: How can coordinated scheduling improve request admission beyond unilateral updates while preserving feasibility and equilibrium properties?}
Shared computing and storage constraints may prevent unilateral improvements from reaching assignments with higher admission cardinality, even when beneficial coordinated adjustments exist. Addressing this limitation requires a scheduling mechanism that combines unilateral and coordinated improvements with stochastic exploration, while ensuring physical feasibility, finite termination, and a well-defined approximate unilateral equilibrium.

To address these questions, we propose Service-Window-Guided Admission and Model Placement (\SWAP), a predictive potential-game approach to continuous UAV edge inference. By incorporating service-window predictions into an admission-priority formulation, \SWAP\ jointly considers wireless service continuity, shared computing resources, and model deployment. A constrained potential game aligns individual scheduling decisions with the system objective, while coordinated admission and finite binary logit refinement extend the search beyond unilateral improvements. The main contributions of this work are summarized as follows.

\begin{itemize}
	\item We formulate a joint request admission, edge-server assignment, and model selection problem for continuous UAV edge inference. The formulation captures the interactions among shared computing resources, model storage, and persistent session-state migration. An admission-priority objective incorporates delay, interruption risk, model quality, migration, and deployment costs as bounded secondary losses, ensuring that any feasible increase in admitted request cardinality strictly improves the objective.
	
	\item We develop a predictive service-window mechanism that incorporates causal mobility and wireless observations into continuity-aware scheduling. Quantile regression and probability calibration characterize the remaining connection duration and interruption risk, while a bounded residence factor captures the potential reuse of shared model instances. These predictions guide residence-aware model deployment and the amortization of model-loading costs, while physical service feasibility is determined by the full execution delay.
	
	\item We establish a constrained exact potential game based on marginal-contribution utilities and develop a scheduling algorithm combining unilateral improvement, model consolidation, two-request admission, and finite binary logit refinement. Theoretical analysis establishes the NP-hardness of the optimization problem, the existence of a constrained pure Nash equilibrium, and the finite termination of the descent procedure. The returned assignment remains feasible, satisfies an approximate unilateral equilibrium condition, and achieves an objective value no greater than that of the initial descent solution.
	
	\item We conduct extensive numerical evaluations under varying request loads, mobility patterns, and edge-resource conditions. Comparisons with five adapted benchmarks, together with prediction assessments, ablation studies, and parameter sensitivity analyses, demonstrate the effectiveness of \SWAP\ in improving service continuity and request admission under resource contention. The results further characterize the associated tradeoffs in model-deployment traffic, checkpoint migration, inference quality, and execution delay, as well as the computational cost of coordinated scheduling.
\end{itemize}

\section{Literature Review}\label{sec:related}

Research on continuous UAV edge inference is closely related to three directions: computation offloading and resource coordination, model deployment and service caching, and mobility-aware service continuity. This section reviews representative studies in these areas, with particular attention to their treatment of shared resources, model availability, and service persistence. Table~\ref{tab:related} summarizes the principal research directions and contributions of the related studies.

\noindent\textit{(i) Offloading and Resource Coordination:}
Computation offloading enables mobile devices to access external computing resources and distribute computational workloads across available execution nodes. Kuang et al.~\cite{utility} investigated utility-aware UAV deployment and task offloading, while Chen et al.~\cite{trajectory} considered joint UAV trajectory optimization and resource allocation. Resource pricing has also been explored to coordinate competing offloading demands~\cite{pricing}. Other studies have incorporated execution deadlines into task scheduling~\cite{deadline} and mobility awareness into node selection in hierarchical edge systems~\cite{hier}. These approaches highlight the importance of coordinating execution placement and resource allocation under heterogeneous service requirements and dynamic network conditions.

Game-theoretic approaches have further been employed to address the interactions among offloading decisions competing for shared edge resources. Zhou et al.~\cite{wcog} studied workload-oriented computation offloading in UAV-assisted mobile edge computing, whereas Wu et al.~\cite{sotnra} investigated joint computation offloading and resource allocation in heterogeneous UAV-assisted networks. Liu et al.~\cite{pgevc} formulated computation offloading in edge--vehicle collaboration as a potential game, while Chai et al.~\cite{sgto} developed a Stackelberg game for joint computing resource allocation and task offloading. These studies provide valuable approaches to coordinating resource competition and interdependent execution decisions. In continuous UAV edge inference, however, request admission and server assignment also affect the deployment and reuse of inference models. Admitting or redirecting a request may change both the computation delay of existing requests and the model instances required at each server. Such interactions motivate joint consideration of request admission and shared model deployment beyond individual offloading costs.

\noindent\textit{(ii) Model Deployment and Service Caching:}
The provision of heterogeneous edge inference services requires suitable inference models to be deployed alongside available computing resources. Different models have distinct computational requirements, storage footprints, and inference-quality characteristics, making model selection and deployment important aspects of edge service provisioning. Huang et al.~\cite{huangdnn} investigated joint deep neural network (DNN) deployment, selection, and configuration for heterogeneous inference services, while Siar et al.~\cite{qos} studied quality-of-service-aware model selection in edge intelligence. Tian et al.~\cite{cpn} considered joint AI model deployment, multi-hop routing, and resource orchestration in UAV-assisted computing-power networks. Related studies have also explored edge-assisted model fetching~\cite{accsi} and joint DNN partitioning, computation offloading, and resource allocation~\cite{dnnveh}, demonstrating the close relationship between model availability and execution-resource management.

Service caching further improves resource utilization by enabling the reuse of deployed models and service components across successive requests. Ham et al.~\cite{twist} investigated integrated service caching and computation offloading in hierarchical cloud architectures. Other studies have explored collaborative service caching and resource allocation~\cite{collabcaching}, learning-based caching in edge--cloud systems~\cite{drlcache}, and cooperative caching under energy-harvesting constraints~\cite{ehcache}. These studies demonstrate the importance of adapting service caching and model availability to evolving demands. For continuous UAV edge inference, the potential reuse of a shared model instance is also influenced by the remaining wireless service windows of its associated requests. Incorporating service-window information into model-deployment planning therefore provides an opportunity to account for prospective model reuse alongside immediate loading and storage costs.

\begin{table*}[t!]
	\centering
	\caption{Summary of representative related studies and the proposed method.}
	\label{tab:related}
	\scriptsize\fontsize{8}{9.3}\selectfont
	\setlength{\tabcolsep}{4.0pt}
	\renewcommand{\arraystretch}{1.08}
	\begin{tabular}{>{\raggedright\arraybackslash}p{1.5cm} >{\raggedright\arraybackslash}p{3.35cm} >{\raggedright\arraybackslash}p{11.8cm}}
		\toprule
		\textbf{Reference} & \textbf{Research Direction} & \textbf{Representative Contributions} \\
		\midrule
		\cite{wcog,pgevc}
		& Game-theoretic computation offloading
		& Developed game-theoretic offloading methods for UAV-assisted and edge--vehicle computing systems, addressing interdependent execution decisions under shared resource competition. \\
		\cite{sotnra,sgto,pricing}
		& Hierarchical resource coordination
		& Investigated joint computation offloading and resource allocation through game-theoretic coordination, Stackelberg interactions, and resource-pricing mechanisms. \\
		\cite{utility,trajectory,deadline,hier}
		& UAV deployment and mobility-aware scheduling
		& Considered task offloading with UAV deployment, trajectory optimization, resource allocation, execution deadlines, and mobility-aware node selection. \\
		\midrule
		\cite{huangdnn,qos}
		& Heterogeneous edge inference
		& Investigated DNN deployment, model selection, and configuration for inference services with heterogeneous computing and quality-of-service requirements. \\
		\cite{twist,collabcaching,drlcache,ehcache}
		& Service caching and computation offloading
		& Coordinated service caching, computation offloading, and resource allocation through hierarchical, collaborative, and learning-based approaches. \\
		\cite{cpn,accsi,dnnveh}
		& Model deployment and resource orchestration
		& Studied AI model deployment with routing and resource orchestration, edge-assisted model fetching, and joint DNN partitioning, offloading, and resource allocation. \\
		\midrule
		\cite{moei,csm}
		& Mobility-aware inference and service migration
		& Investigated model partitioning and service migration for mobile inference, together with cross-edge migration for latency-sensitive rendering services. \\
		\cite{vehicle,rerouting}
		& Mobility-aware and multi-timescale service management
	& Incorporated mobility awareness into service migration and investigated coordinated migration and task rerouting across different time scales. \\
		\cite{vrccs,micro,twin}
		& Learning-based service placement and migration
		& Developed learning-based placement and migration methods for vehicular edge computing, microservices, and edge-assisted digital twin applications. \\
		\midrule
		\mbox{\textbf{\SWAP}}\newline\mbox{(our method)}
		& Service-window-aware scheduling and shared model deployment
		& Integrates calibrated service-window prediction, strict admission priority, and residence-aware shared model deployment with a constrained exact potential game and coordinated scheduling, yielding a feasible approximate unilateral equilibrium. \\
		\bottomrule
	\end{tabular}
\end{table*}

\noindent\textit{(iii) Mobility-Aware Service Continuity:}
Mobility-aware service management aims to sustain service availability as mobile users move across different edge coverage regions. Liu et al.~\cite{moei} investigated mobility-aware edge inference through model partitioning and service migration, while Li et al.~\cite{csm} studied cross-edge service migration for real-time rendering applications. Wang et al.~\cite{vehicle} incorporated spatial--temporal mobility awareness into service migration for connected vehicles, whereas Shi et al.~\cite{rerouting} investigated the coordination of service migration and task rerouting across different time scales. Learning-based service placement and migration have also been studied in vehicular edge computing~\cite{vrccs}, microservice systems~\cite{micro}, and edge-assisted digital twin applications~\cite{twin}.

These studies highlight the importance of adapting service placement and migration to user mobility and changing resource conditions. Continuous UAV edge inference further requires accounting for the persistence of wireless connections throughout successive inference services. Even when an assignment satisfies the current transmission-rate, model-quality, and execution-delay requirements, service completion may be affected by subsequent wireless disconnection. Moreover, persistent session checkpoints introduce dependencies between successive requests, while shared model instances couple the service decisions of multiple UAVs. Motivated by these characteristics, our work integrates calibrated service-window prediction, admission-priority optimization, and residence-aware shared model deployment within a constrained potential-game framework.

\section{System Overview and Problem Formulation}\label{sec:model}

This section presents the system model and problem formulation for continuous UAV edge inference. We first introduce the network architecture and characterize the UAV requests, ground-edge servers, and inference models. We then describe service assignment and resource-dependent feasibility, followed by service-window prediction and residence-aware shared model deployment. Based on these models, we formulate the joint scheduling problem with a strict admission-priority objective.

\subsection{System Overview}\label{sec:overview}

\begin{figure*}[t]
	\centering
	\includegraphics[width=\textwidth]{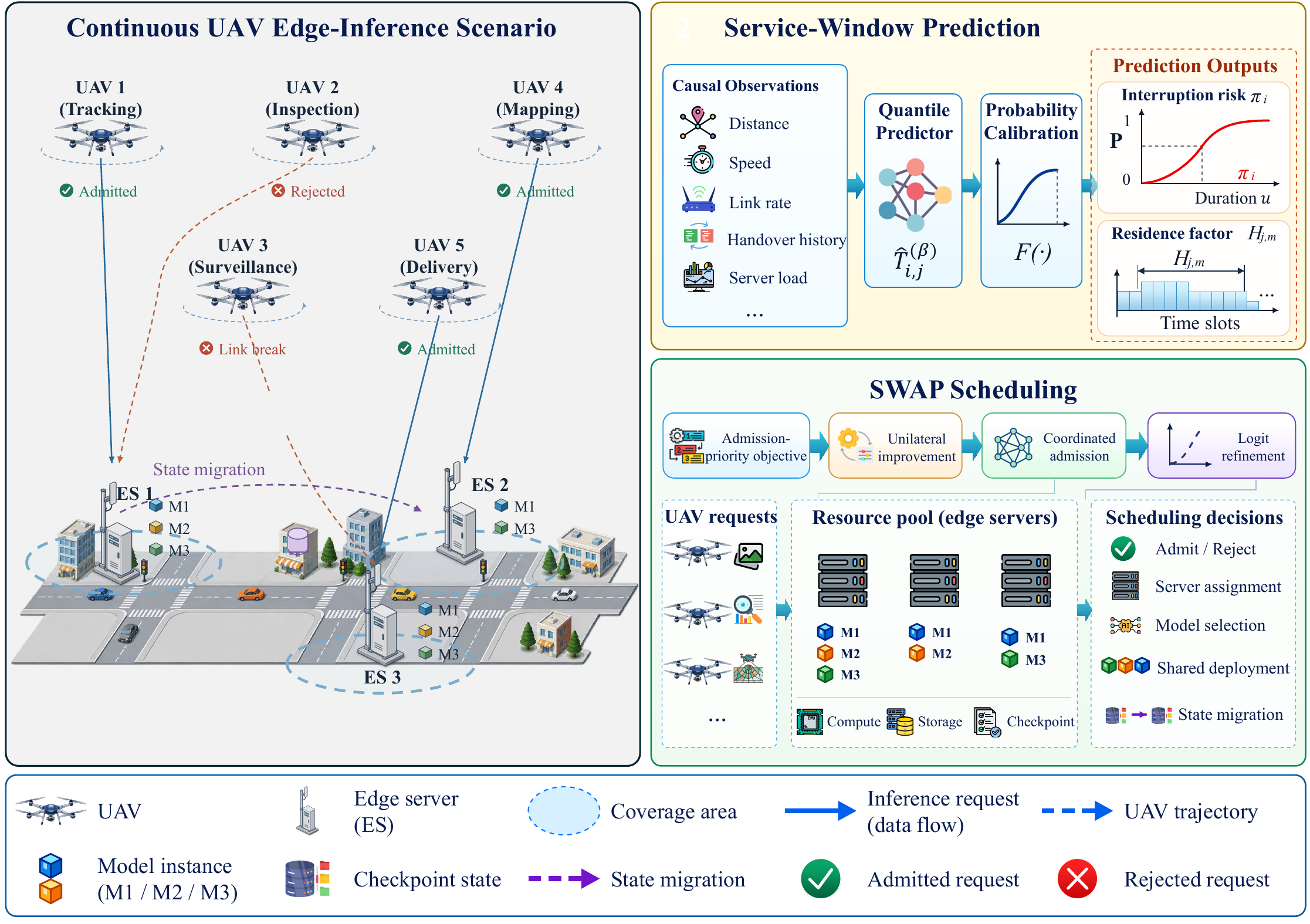}
\caption{System architecture and scheduling framework for continuous UAV edge inference. Service-window quantiles, calibrated interruption risks, and shared model residence factors guide joint request admission, edge-server assignment, and model deployment under physical service constraints.}
	\label{fig:architecture}\vspace{-2mm}
\end{figure*}

As illustrated in Fig.~\ref{fig:architecture}, we consider a continuous UAV edge-inference system comprising mobile UAVs, geographically distributed ground-edge servers, and heterogeneous inference models. UAVs generate inference requests that are processed by ground-edge servers equipped with computing and model-storage resources. Let $\mathcal N=\{n_i\}_{i=1}^{N}$, $\mathcal J=\{e_j\}_{j=1}^{J}$, and $\mathcal M=\{\mathsf{M}_m\}_{m=1}^{M}$ denote the sets of UAVs, ground-edge servers, and inference models, respectively, where $N$, $J$, and $M$ are their corresponding cardinalities. For any positive integer $K$, define $[K]=\{1,\ldots,K\}$, such that $i\in[N]$, $j\in[J]$, and $m\in[M]$ serve as the respective indices.

The system operates over discrete decision slots indexed by $t$, each with duration $\Delta>0$ seconds. Each UAV generates one indivisible inference request per slot and maintains a persistent inference session, while its trajectory evolves independently of scheduling decisions. Multiple UAVs may share a resident model instance at the same server, enabling model reuse and shared computing. The serving edge and selected inference model may change between successive requests, with inter-edge session-state migration required when the serving server differs from the location of the retained checkpoint. At the beginning of each slot, the scheduler observes the current requests, wireless link conditions, server resources, and inherited service states. Based on these observations and the predicted service windows, it jointly determines request admission, edge-server assignment, and model selection. The resulting decisions govern computing resource sharing, model deployment, and session-state migration, subject to the physical service constraints.

\subsection{Modeling of UAVs, Edge Servers, and Inference Models}\label{sec:entities}

\noindent\textit{(i) UAV Requests:}
Each UAV $n_i\in\mathcal N$ generates one indivisible inference request per slot. The request is characterized by an input size $b_i>0$ bits, a nominal computational intensity $\phi_i>0$ central processing unit (CPU) cycles per bit, a service deadline $d_i>0$ seconds, and a dimensionless minimum inference-quality requirement $\underline q_i\in[0,1]$. Each UAV also maintains a persistent inference session with a checkpoint of size $v_i\ge0$ megabytes (MB), where $1$ MB equals $10^6$ bytes. Request attributes are observed at the beginning of each slot, while the session checkpoint is retained across successive requests, including those that are rejected.

\noindent\textit{(ii) Ground-Edge Servers:}
Each ground-edge server $e_j\in\mathcal J$ is equipped with computing capacity $C_j>0$ CPU cycles/s and model-storage capacity $S_j>0$ MB. The computing capacity is shared among admitted requests assigned to the server, while requests selecting the same inference model can share a single resident model instance. Under on-demand model deployment, the resident model set is determined by the admitted requests in each slot. Inter-edge backhaul links enable the migration of persistent session checkpoints when the serving server changes.

\noindent\textit{(iii) Inference Models:}
Each inference model $\mathsf{M}_m\in\mathcal M$ is characterized by a dimensionless inference quality $q_m\in[0,1]$, a workload multiplier $\gamma_m>0$, a storage requirement $s_m>0$ MB, and a cold-start time $c_m\ge0$ seconds. These parameters represent different computation--quality tradeoffs among model variants~\cite{huangdnn,qos}. The models provide compatible inference services and share a common session-state representation. Their quality and resource profiles are specified synthetically in the numerical evaluation.

For request $i$ executed using model $\mathsf{M}_m$, the corresponding computation workload is given by
\begin{equation}
	L_{i,m}=b_i\phi_i\gamma_m.
	\label{eq:workload}
\end{equation}
Here, $L_{i,m}$ is measured in CPU cycles, where $b_i\phi_i$ represents the nominal computation workload and $\gamma_m$ characterizes the computational requirements of the selected inference model.

\subsection{Service Assignment and Physical Feasibility}\label{sec:execution}

For a given decision slot, we omit the time index for simplicity. Let $a_i\in\{0\}\cup([J]\times[M])$ denote the scheduling action of request $i$, where $a_i=0$ represents rejection and $a_i=(j,m)$ assigns the request to server $e_j$ using model $\mathsf{M}_m$. Define the binary assignment indicator $x_{i,j,m}=\ind\{a_i=(j,m)\}$, where $\ind\{\cdot\}$ equals one if the condition holds and zero otherwise. The admission indicator is $r_i=\sum_{j=1}^{J}\sum_{m=1}^{M}x_{i,j,m}\in\{0,1\}$. Accordingly, the joint action profile is $\boldsymbol a=(a_1,\ldots,a_N)$, with each request assigned to at most one server--model pair.

Under the joint assignment $\boldsymbol a$, the computation workload of request $i$ is
\begin{equation}
	L_i=\sum_{j=1}^{J}\sum_{m=1}^{M}x_{i,j,m}L_{i,m}.
	\label{eq:assignedworkload}
\end{equation}
Similarly, the selected model quality is $q_i=\sum_{j=1}^{J}\sum_{m=1}^{M}x_{i,j,m}q_m$. For an admitted request with $a_i=(j,m)$, we have $L_i=L_{i,m}>0$ and $q_i=q_m$. Both quantities depend on the joint assignment, while rejected requests contribute neither computation workload nor model-quality loss.

\noindent\textit{Shared Model Placement:}
Under on-demand model deployment, a model instance is resident at server $e_j$ if at least one admitted request uses it. Accordingly, the binary placement indicator is
\begin{equation}
	y_{j,m}=\ind\!\left\{\sum_{i=1}^{N}x_{i,j,m}>0\right\}.
	\label{eq:placement}
\end{equation}
Let $y^-_{j,m}\in\{0,1\}$ indicate whether model $\mathsf{M}_m$ is already resident at server $e_j$ before the current decision. A newly loaded instance is identified by $z_{j,m}=y_{j,m}(1-y^-_{j,m})$. The deployed models must satisfy the storage constraint
\begin{equation}
	\sum_{m=1}^{M}s_my_{j,m}\le S_j,\qquad j\in[J].
	\label{eq:storage}
\end{equation}
Requests assigned to the same server--model pair share one model instance, whose storage requirement is counted only once. Unused model instances are evicted before new instances are loaded, with transient storage overlap and eviction latency excluded from the model. Session checkpoints are maintained independently of model residency and persist across rejected requests.

\noindent\textit{Wireless Transmission:}
For an admitted request assigned to server $e_j$, the effective uplink rate $R_{i,j}$ is modeled using the Shannon expression for an interference-free wireless link~\cite{tse2005fundamentals}:
\begin{equation}
	R_{i,j}=B\log_2\!\left(1+\frac{Pg_{i,j}}{N_0B}\right).
	\label{eq:rate}
\end{equation}
Here, $B>0$ is the effective per-link bandwidth in Hz, $P>0$ denotes the UAV transmission power in W, $g_{i,j}\ge0$ is the dimensionless channel power gain, and $N_0>0$ is the noise power spectral density in W/Hz. The corresponding upload delay is $D_i^{\rm tx}=b_i/R_{i,j}$. A server is accessible to UAV $i$ when $R_{i,j}\ge R_{\min}$, where $R_{\min}>0$ is the minimum required uplink rate. We consider a fixed effective bandwidth allocation for each link and focus on uplink transmission. The observed rate determines the upload delay, while subsequent channel variations affect wireless connection persistence during service execution.

\noindent\textit{Computing Resource Sharing:}
The computation delay of an admitted request depends on its workload and the computing resources shared with other requests at the same server. Let $\Sset_j=\{i\in[N]\mid a_i=(j,m)\text{ for some }m\in[M]\}$ denote the set of admitted-request indices assigned to server $e_j$. For a given assignment, each request $i\in\Sset_j$ has workload $L_i>0$ and receives computing rate $f_i>0$ in CPU cycles/s. We consider the allocation that minimizes the aggregate computation delay $\sum_{i\in\Sset_j}L_i/f_i$, subject to $\sum_{i\in\Sset_j}f_i\le C_j$.
By the Cauchy--Schwarz inequality, the aggregate computation delay satisfies
\begin{equation}
	\sum_{i\in\Sset_j}\frac{L_i}{f_i}
	\ge
	\frac{\left(\sum_{i\in\Sset_j}\sqrt{L_i}\right)^2}{C_j}.
	\label{eq:cpubound}
\end{equation}
For a nonempty $\Sset_j$, equality holds when the server capacity is fully allocated and $f_i$ is proportional to $\sqrt{L_i}$. The optimal computing allocation is therefore
\begin{equation}
	f_i=\frac{C_j\sqrt{L_i}}
	{\sum_{k\in\Sset_j}\sqrt{L_k}},
	\qquad i\in\Sset_j.
	\label{eq:cpu}
\end{equation}
Substituting~\eqref{eq:cpu} into $D_i^{\rm cmp}=L_i/f_i$ gives
\begin{equation}
	D_i^{\rm cmp}
	=\frac{\sqrt{L_i}}{C_j}
	\sum_{k\in\Sset_j}\sqrt{L_k}.
	\label{eq:compute}
\end{equation}
This allocation minimizes the aggregate computation delay for the given assignment, while individual service feasibility is determined by the resulting request delays.

\noindent\textit{Session-State Migration and Model Loading:}
Continuous inference requires both the selected model and the corresponding session state to be available at the serving edge~\cite{moei,csm}. Let $\ell_i^-\in\{-1\}\cup[J]$ denote the index of the server holding the latest checkpoint of UAV $i$'s persistent session, with $\ell_i^-=-1$ before its first admission. For an admitted action $a_i=(j,m)$, the migration indicator is $\mu_i=\ind\{\ell_i^-\ge0,\ j\ne\ell_i^-\}$.

Given an effective inter-edge backhaul rate $R_{\rm bh}>0$ bits/s, the checkpoint-migration delay is $D_i^{\rm mig}=\mu_i 8\times10^6v_i/R_{\rm bh}$, where the factor $8\times10^6$ converts the checkpoint size from MB to bits. Similarly, a model not previously resident at the selected server incurs cold-start delay $D_i^{\rm cold}=c_m(1-y^-_{j,m})$. The cold-start time includes model loading and initialization, while session migration is accounted for separately. Each newly deployed model instance is loaded once, and its associated requests experience the corresponding model-readiness delay.

Combining wireless transmission, computation, session-state migration, and model initialization, the physical service delay of an admitted request is
\begin{equation}
	D_i=D_i^{\rm tx}+D_i^{\rm cmp}
	+D_i^{\rm mig}+D_i^{\rm cold}.
	\label{eq:delay}
\end{equation}

\noindent\textit{Physical Service Feasibility:}
Each admitted action $a_i=(j,m)$ must satisfy the uplink-rate, model-quality, and service-deadline requirements:
\begin{equation}
	R_{i,j}\ge R_{\min},\quad
	q_m\ge\underline q_i,\quad
	D_i\le d_i.
	\label{eq:feasible}
\end{equation}
Together with the storage constraint in~\eqref{eq:storage} and the single-assignment requirement, these conditions define the feasible joint-action set $\A_t$. Since the computation delay in~\eqref{eq:compute} depends on all requests sharing the same server, feasibility is evaluated for the entire joint assignment.

Session checkpoints persist across rejected slots and are updated when the corresponding UAV is admitted. We assume that the latest checkpoint remains available following wireless interruptions and that inference sessions persist throughout each episode. Thus, subsequent admission to a different server requires checkpoint migration. Scheduling decisions take effect at slot boundaries, and scheduling computation is excluded from the physical execution timeline. With $d_i<\Delta$, each request must complete within its current slot, while the associated session may continue across successive slots.

For an accessible UAV--server link $(i,j)$, let $T_{i,j}\ge0$ denote the remaining connection duration until its uplink rate first falls below $R_{\min}$. Successful execution requires the serving wireless connection to remain available throughout the service interval, i.e., $T_{i,j}\ge D_i$. However, the remaining connection duration is unknown at the decision instant. This motivates the service-window prediction mechanism introduced in the following subsection, which incorporates anticipated wireless connectivity into scheduling decisions.

\subsection{Service-Window Prediction and Shared Deployment}\label{sec:predictionmodel}

To characterize uncertainty in future wireless connectivity, we predict the remaining service window using causal mobility and channel observations. For each UAV--server pair $(i,j)$, the feature vector $h_{i,j}$ includes the current link distance and its recent slope, radial velocity, UAV speed, two heading components, current uplink rate and its recent slope, recent handovers, and previous server load. Temporal slopes are estimated from at most ten observations available at the decision instant, allowing predictions to be obtained from historical and current system information.

\noindent\textit{Service-Window Prediction and Interruption Risk:}
Let $T_{\max}>0$ denote the maximum prediction horizon in seconds. Each training label $T_n$ represents the realized service window capped at $T_{\max}$, i.e., $T_n=\min(T_{i,j},T_{\max})$ for the corresponding sample. We employ three quantile gradient-boosted decision tree (Quantile-GBDT) regressors to estimate the quantiles $\beta\in\{0.1,0.5,0.9\}$. For each UAV--server pair, the predicted quantile is $\widehat T_{i,j}^{(\beta)}=\widehat T_\beta(h_{i,j})$. Following quantile regression~\cite{quantile}, each regressor minimizes the pinball loss
\begin{equation}
	\mathcal L_\beta
	=\sum_{n=1}^{n_{\rm tr}}
	\rho_\beta\!\left(T_n-\widehat T_\beta(h_n)\right).
	\label{eq:pinball}
\end{equation}
Here, $n_{\rm tr}$ is the number of training samples, $h_n$ is the feature vector of sample $n$, and $\rho_\beta(e)=e(\beta-\ind\{e<0\})$ is the quantile check-loss function, with $e=T_n-\widehat T_\beta(h_n)$ denoting the prediction residual. The regressors are implemented using LightGBM~\cite{lightgbm}. Their predictions are restricted to the interval $[0,T_{\max}]$ and ordered with a strictly positive separation to prevent quantile crossing.

The predicted quantiles are further used to estimate the probability of wireless interruption within a given duration. For a queried duration $u\in[0,T_{\max}]$, we construct a piecewise-linear approximation $\widetilde F_{i,j}(u)$ by interpolating the endpoints $(0,0)$ and $(T_{\max},1)$ together with the quantile knots $(\widehat T_{i,j}^{(\beta)},\beta)$ for $\beta\in\{0.1,0.5,0.9\}$. Since service-window observations are sampled on a discrete temporal grid, we consider the interruption event $T_{i,j}<u$. To improve the agreement between predicted probabilities and observed interruption frequencies, an isotonic calibration map $\mathcal I$ is fitted using separate calibration samples and binary targets $\ind\{T_{i,j}<u\}$ at predefined durations
~\cite{niculescu2005probabilities}.
The calibrated interruption-probability function is $\widehat F_{i,j}(u)=\mathcal I(\widetilde F_{i,j}(u))$, where $\mathcal I:[0,1]\to[0,1]$ is nondecreasing with endpoint probabilities fixed at zero and one. We further set $\widehat F_{i,j}(u)=0$ for $u\le0$ and $\widehat F_{i,j}(u)=1$ for $u\ge T_{\max}$. For an admitted request under joint assignment $\boldsymbol a$, the predicted interruption risk is defined as
\begin{equation}
	\pi_i(\boldsymbol a)
	=\widehat F_{i,j}\!\left(D_i(\boldsymbol a)+\delta\right).
	\label{eq:risk}
\end{equation}
Here, $\delta\ge0$ is a look-ahead interval in seconds, and $D_i(\boldsymbol a)$ denotes the physical service delay under the current assignment. While $\widehat F_{i,j}(D_i)$ estimates the probability of interruption during service execution, $\pi_i$ additionally accounts for potential connection loss over the look-ahead interval. The resulting risk estimate is incorporated into the scheduling objective to favor assignments with greater expected service continuity.

\noindent\textit{Residence-Aware Shared Deployment:}
Service-window predictions also provide information about the potential reuse of shared model instances. Since each UAV generates successive requests within a persistent inference session, a longer remaining connection duration may support repeated use of the same model across subsequent slots. Accordingly, we define a bounded residence factor for each admitted request $i$ as
\begin{equation}
	H_i(\boldsymbol a)
	=\clip\!\left(
	\frac{\widehat T_{i,j}^{(0.1)}-D_i(\boldsymbol a)}
	{\Delta},1,H\right).
	\label{eq:residence}
\end{equation}
Here, $\widehat T_{i,j}^{(0.1)}$ is the predicted lower quantile of the remaining service window, $H\ge1$ denotes the maximum residence factor, and $\clip(x,l,u)=\min\{u,\max\{l,x\}\}$. The factor $H_i\in[1,H]$ represents a continuous estimate of potential model reuse over successive slots.

For a model instance $\mathsf{M}_m$ shared by requests at server $e_j$, the corresponding residence factor is
\begin{equation}
	H_{j,m}(\boldsymbol a)=
	\begin{cases}
		\displaystyle\max_{i\in[N]\mid x_{i,j,m}=1}
		H_i(\boldsymbol a), & y_{j,m}=1,\\
		1, & y_{j,m}=0.
	\end{cases}
	\label{eq:grouphorizon}
\end{equation}
Taking the maximum accounts for the possibility that a shared model instance remains useful to a persistent session after other associated connections have ended. The resulting factor characterizes the anticipated reuse of each shared model instance and guides residence-aware deployment decisions. To incorporate potential model reuse into scheduling, we amortize the cold-start delay over the shared residence factor and define the planning delay as
\begin{equation}
	\widetilde D_i
	=D_i^{\rm tx}+D_i^{\rm cmp}+D_i^{\rm mig}
	+\frac{D_i^{\rm cold}}{H_{j,m}}.
	\label{eq:planningdelay}
\end{equation}
The migration delay remains an immediate cost, whereas model initialization is amortized according to the estimated reuse opportunity. Since $H_{j,m}\ge1$, the planning delay satisfies $0\le\widetilde D_i\le D_i$. The planning delay guides scheduling preferences, while service feasibility continues to be determined by the full physical delay $D_i$ in~\eqref{eq:delay}.

\subsection{Problem Formulation}\label{sec:formulation}

Based on the preceding system models, we formulate the joint request admission, edge-server assignment, and model selection problem for continuous UAV edge inference. The objective is to prioritize request admission while accounting for service delay, wireless interruption risk, inference quality, and reconfiguration costs. To this end, we introduce five normalized secondary losses for a feasible joint action profile $\boldsymbol a\in\A_t$. Let $q_{\max}=\max_{m\in[M]}q_m$ and $s_{\max}=\max_{m\in[M]}s_m$, and choose $v_{\max}>0$ such that $v_i\le v_{\max}$ for all $i\in[N]$. We assume $\underline q_i<q_{\max}$ for each request. The dependence of individual losses on $\boldsymbol a$ is omitted for brevity.

First, the planning-delay loss and predicted interruption-risk loss are defined as
\begin{equation}
	\begin{aligned}
		G_D=\frac{1}{N}\sum_{i\in[N]\mid r_i=1}
		\frac{\widetilde D_i}{d_i},~~~~
		G_R=\frac{1}{N}\sum_{i\in[N]\mid r_i=1}\pi_i.
	\end{aligned}
	\label{eq:lossrisk}
\end{equation}
Here, $G_D$ measures the deadline-normalized planning delay, while $G_R$ captures the predicted wireless interruption risk. The inference-quality loss is given by
\begin{equation}
	G_Q=\frac{1}{N}\sum_{i\in[N]\mid r_i=1}
	\left(\frac{q_{\max}-q_i}
	{q_{\max}-\underline q_i}\right)^p,
	\label{eq:lossq}
\end{equation}
where $p\ge1$ controls the preference among models satisfying the minimum inference-quality requirement.

We further consider the costs associated with session-state migration and shared model deployment. The migration loss depends on the transferred checkpoint size, whereas the deployment loss accounts for newly loaded model instances and their estimated residence factors:
\begin{equation}
	{\small\begin{aligned}
		G_M=\frac{1}{Nv_{\max}}
		\sum_{i\in[N]\mid r_i=1}\mu_i v_i,~~~
		G_P=\frac{1}{Ns_{\max}}
		\sum_{j=1}^{J}\sum_{m=1}^{M}
		\frac{z_{j,m}s_m}{H_{j,m}}.
	\end{aligned}}
	\label{eq:lossmp}
\end{equation}
Each newly loaded server--model instance incurs one deployment cost, which is amortized according to its estimated shared residence factor.

The five secondary losses are dimensionless and satisfy $0\le G_k\le1$ for all $k\in\mathcal K=\{D,R,Q,M,P\}$ on the feasible set. The bounds on planning delay, interruption risk, model quality, and checkpoint size ensure the boundedness of the first four losses. Moreover, under on-demand model placement, the number of resident model instances cannot exceed the number of admitted requests. Together with $H_{j,m}\ge1$ and $s_m\le s_{\max}$, this also bounds the deployment loss.

With nonnegative weights $w_k$ satisfying $\sum_{k\in\mathcal K}w_k=1$, the aggregate secondary loss is $G(\boldsymbol a)=\sum_{k\in\mathcal K}w_kG_k$, where $0\le G(\boldsymbol a)\le1$. We define the admission-priority objective as
\begin{equation}
	F_t(\boldsymbol a)
	=\sum_{i=1}^{N}(1-r_i)+\eta G(\boldsymbol a),
	\label{eq:admissioncost}
\end{equation}
where $0<\eta<1$ determines the contribution of secondary performance losses. Accordingly, the per-slot joint scheduling problem is formulated as
\begin{equation}
	(\mathcal P_t):\quad
	\min_{\boldsymbol a} F_t(\boldsymbol a).
	\label{eq:objective}
\end{equation}

\noindent\textbf{s.t.}

\noindent\textbf{\textit{\underline{Constraints}:}}
\begin{itemize}
	\item \textit{Unique request assignment:}
	$a_i\in\{0\}\cup([J]\times[M])$ and $r_i=\sum_{j=1}^{J}\sum_{m=1}^{M}x_{i,j,m}\in\{0,1\}$, for all $i\in[N]$.
	
	\item \textit{Shared model placement and storage capacity:}
	\eqref{eq:placement} and \eqref{eq:storage}, with $z_{j,m}=y_{j,m}(1-y^-_{j,m})$.
	
	\item \textit{Computing resource allocation:}
	\eqref{eq:cpu} for all admitted requests.
	
	\item \textit{Physical service feasibility:}
	\eqref{eq:feasible} for every admitted server--model assignment.
\end{itemize}

\noindent\textbf{\textit{\underline{Variables}:}}
\begin{itemize}
	\item \textit{Joint scheduling decisions:}
	$\boldsymbol a=(a_1,\ldots,a_N)$, where $a_i=0$ denotes rejection and $a_i=(j,m)$ specifies the selected server--model pair.
	
	\item \textit{Induced admission and deployment variables:}
	$\{r_i,x_{i,j,m}\}$ and $\{y_{j,m},z_{j,m}\}$, determined by the joint scheduling decisions and inherited model states.
\end{itemize}

\noindent 
In $\mathcal P_t$, the joint action profile $\boldsymbol a$ determines request admission, edge-server assignment, and model selection, which further induce model placement, loading decisions, and computing resource allocation. The feasible set $\A_t$ incorporates the single-assignment requirement, model-storage capacity, and physical service constraints. Since multiple requests share computing resources and model instances, changing one request's assignment may affect the computation delay and deployment costs of other requests. The resulting optimization therefore accounts for these interdependent decisions at the system level, with request attributes, resource states, and service-window predictions fixed within each decision slot.

The objective $F_t$ establishes a strict priority for request admission while incorporating delay, interruption risk, model quality, migration, and deployment costs as secondary considerations. Since $0\le G(\boldsymbol a)\le1$ and $0<\eta<1$, any feasible increase in admitted request cardinality strictly decreases $F_t$, while the weighted secondary losses distinguish assignments admitting the same number of requests. This property is formally established in Proposition~\ref{prop:admission}. However, the coupling among discrete scheduling decisions, shared computing resources, and model deployment renders $\mathcal P_t$ NP-hard, as established in Proposition~\ref{prop:hardness}. To address this challenge, we develop a constrained potential-game approach in the following section to coordinate individual scheduling decisions toward the common admission-priority objective.

\section{Design Methodology of SWAP}\label{sec:solution}

This section develops \SWAP\ for the admission-priority scheduling problem $	\mathcal P_t$. We first construct a constrained potential game to coordinate interdependent service decisions, and then enhance unilateral improvement with coordinated admission and finite binary logit refinement. 

\subsection{Constrained Potential-Game Formulation}\label{sec:game}

At each decision slot, request attributes, wireless link rates, server capacities, inherited service states, and service-window predictions remain fixed during optimization. However, individual scheduling decisions are coupled through shared computing resources and model deployment. Changing the assignment of one request may affect the computation delay and model-deployment costs of other requests. We therefore formulate the scheduling problem as a constrained game in which each request acts as a player, and its utility reflects its contribution to the common admission-priority objective.

For each player $i\in[N]$, the action alphabet is defined as
\begin{equation}
	\mathcal B_i=\{0\}\cup
	\{(j,m)\in[J]\times[M]\mid
	R_{i,j}\ge R_{\min},\ q_m\ge\underline q_i\}.
	\label{eq:alphabet}
\end{equation}
Here, $a_i=0$ denotes request rejection, while $a_i=(j,m)$ represents a candidate server--model assignment satisfying the current uplink-rate and inference-quality requirements. Given the actions $\boldsymbol a_{-i}$ of all other players, the feasible strategy set of player $i$ is
\begin{equation}
	\A_i(\boldsymbol a_{-i})
	=\{\bar a_i\in\mathcal B_i\mid
	(\bar a_i,\boldsymbol a_{-i})\in\A_t\}.
	\label{eq:deviation}
\end{equation}
Here, $\bar a_i$ denotes a candidate action, and $(\bar a_i,\boldsymbol a_{-i})$ is the resulting joint profile. Unlike the action alphabet $\mathcal B_i$, the feasible set $\A_i(\boldsymbol a_{-i})$ additionally accounts for shared model storage and computation-dependent service deadlines. Thus, each player's feasible strategies depend on the assignments of other requests.

To align individual improvements with the system objective, we define the marginal-contribution utility of player $i$ as
\begin{equation}
	u_i(\boldsymbol a)
	=F_t(0,\boldsymbol a_{-i})-F_t(\boldsymbol a).
	\label{eq:utility}
\end{equation}
Here, $F_t(\boldsymbol a)$ is the admission-priority objective in~\eqref{eq:admissioncost}, and $F_t(0,\boldsymbol a_{-i})$ represents the system cost when request $i$ is rejected while all other actions remain unchanged. The latter is evaluated with the corresponding changes in computing resource sharing, model occupancy, and residence-dependent costs. As established in Lemma~\ref{lem:downward}, removing an admitted request preserves physical feasibility. The resulting utility therefore captures the contribution of each request to the overall scheduling objective, including its effects on other requests.
Consider a feasible unilateral deviation from $\boldsymbol a$ to $\boldsymbol b=(\bar a_i,\boldsymbol a_{-i})$. Since the actions of all other players remain unchanged, the counterfactual term $F_t(0,\boldsymbol a_{-i})$ is identical for both profiles. Consequently, $u_i(\boldsymbol b)-u_i(\boldsymbol a)=F_t(\boldsymbol a)-F_t(\boldsymbol b)$. It follows that $\Phi_t(\boldsymbol a)=-F_t(\boldsymbol a)$ constitutes an exact potential function~\cite{potential}, and every feasible unilateral utility improvement produces an equivalent decrease in the system objective.
Based on this property, unilateral descent selects the feasible action update that yields the largest decrease in $F_t$. An update is accepted when the objective improvement exceeds a prescribed tolerance $\eps$, where $0<\eps<1-\eta$. Since any feasible increase in admitted request cardinality decreases $F_t$ by at least $1-\eta$, the improvement rule preserves the strict admission priority established in Section~\ref{sec:formulation}.

\subsection{Coordinated Admission and Binary Logit Refinement}\label{sec:coordination}

Although unilateral descent aligns individual improvements with the system objective, it may terminate at a constrained local equilibrium that admits fewer requests than another feasible assignment. This limitation arises when admitting additional requests requires simultaneous adjustments to existing server assignments or model selections. To explore improvements beyond unilateral deviations, \SWAP\ introduces two coordinated admission mechanisms: model consolidation and two-request admission. Each coordinated adjustment must satisfy the physical service constraints and is accepted only if it decreases $F_t$ by more than $\eps$.

\noindent\textit{Model Consolidation:}
For each server $e_j$ and candidate model $\mathsf{M}_m$, the coordinator examines whether the requests currently assigned to that server can share the candidate model while satisfying their inference-quality and full service-delay requirements. When consolidation is feasible, previously rejected requests are considered for admission in increasing order of $\sqrt{L_{i,m}}/d_i$, where $L_{i,m}$ denotes the computation workload under the candidate model. Among at most $JM$ server--model candidates, the adjustment yielding the largest admissible decrease in $F_t$ is selected. By consolidating multiple requests onto a shared model instance, this mechanism can reduce storage occupancy and create additional admission opportunities.

\noindent\textit{Two-Request Admission:}
When model consolidation yields no sufficient improvement, the coordinator considers joint adjustments involving one rejected request and at most one admitted request. The rejected request is assigned to a feasible server--model pair, while the admitted request may change its serving server or selected model without losing admission. These changes are evaluated jointly, allowing coordinated reassignment to accommodate additional requests under shared resource constraints. Following each accepted adjustment, unilateral descent is resumed. The coordinated descent procedure terminates when no considered unilateral or coordinated move decreases $F_t$ by more than $\eps$.

To further explore the feasible assignment space, we introduce finite binary logit refinement~\cite{loglinear}. Starting from the coordinated descent solution, each proposal randomly selects a player $i$ uniformly from $[N]$ and a candidate action $\bar a_i$ uniformly from its action alphabet $\mathcal B_i$. The proposed profile is $\boldsymbol b=(\bar a_i,\boldsymbol a_{-i})$. Proposals outside $\A_t$ are rejected, while feasible proposals are accepted with probability
\begin{equation}
	p_k^{\rm acc}(\boldsymbol a,\boldsymbol b)
	=\frac{1}{1+\exp\!\left(
		\frac{F_t(\boldsymbol b)-F_t(\boldsymbol a)}{\tau_k}
		\right)}.
	\label{eq:logit}
\end{equation}
Here, $p_k^{\rm acc}(\boldsymbol a,\boldsymbol b)$ denotes the acceptance probability at proposal $k$, and $\tau_k>0$ is the corresponding temperature. Proposals with lower objective values are more likely to be accepted, while the nonzero acceptance probability of higher-cost moves enables exploration of alternative feasible assignments.
The temperature decreases according to the finite geometric cooling schedule
\begin{equation}
	\tau_k=\max\!\left\{
	10^{-12},\frac{\eta c_\tau}{N}\alpha^k
	\right\},\quad k=0,\ldots,K_{\mathrm L}-1.
	\label{eq:temperature}
\end{equation}
Here, $K_{\mathrm L}$ is the number of proposals, set to $5N$ by default; $c_\tau>0$ controls the temperature scale, and $\alpha\in(0,1)$ is the cooling factor. The coefficient $\eta$ follows the admission-priority objective in~\eqref{eq:admissioncost}, while the positive lower bound maintains a nonzero temperature throughout the refinement.

Since the action alphabet $\mathcal B_i$ remains fixed within each slot, two profiles differing only in player $i$'s action have the same proposal probability $1/(N|\mathcal B_i|)$ in either direction. For a fixed positive temperature $\tau$, this symmetric proposal rule and the binary logit acceptance probability satisfy detailed balance with a distribution proportional to $\exp(-F_t/\tau)$. Together with the connectivity of the feasible-profile graph established in Lemma~\ref{lem:downward}, this property characterizes the preference of the corresponding fixed-temperature process for lower-cost feasible assignments. The finite cooling schedule uses this preference to guide stochastic exploration beyond deterministic descent.

During refinement, the algorithm retains both the best profile visited and the final profile of the logit trajectory. Each distinct candidate subsequently undergoes coordinated descent, and the incumbent is updated only when the refined candidate decreases $F_t$ by more than $\eps$. In this way, stochastic exploration is combined with coordinated improvement to obtain a feasible assignment whose objective value is no greater than that of the initial descent solution.

\subsection{Overall Procedure}\label{sec:algorithm}

Algorithm~\ref{alg:airco} summarizes the overall procedure of \SWAP. Each decision slot involves service-window prediction, feasible initialization, coordinated potential descent, and finite binary logit refinement. Here, $\operatorname{Descend}$ denotes the iterative procedure combining feasible unilateral improvement, model consolidation, and two-request admission, where each accepted update reduces $F_t$ by more than $\eps$.

\par\addvspace{6pt}\noindent
\begin{minipage}{\columnwidth}
	\begin{algorithm}[H]
		\scriptsize\setstretch{0.99}
		\caption{Proposed \SWAP\ for one decision slot}
		\label{alg:airco}
		\LinesNumbered
		\SetAlFnt{\scriptsize}
		\SetAlCapFnt{\small\bfseries}
		\SetAlCapNameFnt{\small\bfseries}
		\SetKwInput{Input}{Input}
		\SetKwInput{Output}{Return}
		\Input{Current requests, link conditions, server resources, inherited states, causal histories, $K_{\mathrm L}$, and $\eps$.}
		Predict service-window quantiles and interruption risks\;\nllabel{line:forecast}
		Construct action sets $\mathcal B_i$ using~\eqref{eq:alphabet}\;
		Initialize a feasible profile $\boldsymbol a^0\in\A_t$\;\nllabel{line:repair}
		$\boldsymbol a^L\leftarrow\operatorname{Descend}(\boldsymbol a^0)$\;\nllabel{line:descent}
		$\boldsymbol a,\boldsymbol a^B\leftarrow\boldsymbol a^L,\boldsymbol a^L$\;
		\For{$k=0$ \KwTo $K_{\mathrm L}-1$}{
			Sample $i\in[N]$ and $\bar a_i\in\mathcal B_i$ uniformly\;
			$\boldsymbol b\leftarrow(\bar a_i,\boldsymbol a_{-i})$\;
			\If{$\boldsymbol b\in\A_t$}{
				Draw $\xi\sim\operatorname{Uniform}(0,1)$\;
				\If{$\xi<p_k^{\rm acc}(\boldsymbol a,\boldsymbol b)$}{
					$\boldsymbol a\leftarrow\boldsymbol b$\;
					\If{$F_t(\boldsymbol a)<F_t(\boldsymbol a^B)$}{
						$\boldsymbol a^B\leftarrow\boldsymbol a$\;
					}
				}
			}
		}\nllabel{line:walkend}
		$\boldsymbol a^\star\leftarrow\boldsymbol a^L$\;\nllabel{line:polish}
		\For{each distinct $\boldsymbol b\in\{\boldsymbol a^B,\boldsymbol a\}\setminus\{\boldsymbol a^L\}$}{
			$\boldsymbol c\leftarrow\operatorname{Descend}(\boldsymbol b)$\;
			\If{$F_t(\boldsymbol c)<F_t(\boldsymbol a^\star)-\eps$}{
				$\boldsymbol a^\star\leftarrow\boldsymbol c$\;
			}
		}
		Verify $\boldsymbol a^\star\in\A_t$ and compute $\mathcal E(\boldsymbol a^\star)$\;\nllabel{line:certificate}
		\Output{Feasible assignment $\boldsymbol a^\star$ and unilateral gap $\mathcal E(\boldsymbol a^\star)$.}
	\end{algorithm}
\end{minipage}
\par\addvspace{7pt}

\noindent\textbf{Step 1. Prediction and Feasible Initialization:}
Service-window predictions and feasible action sets are obtained from causal observations. The inherited assignment is adjusted by removing rate- or quality-infeasible actions and retaining storage-feasible model subsets that preserve the most requests. Remaining violations are addressed through feasible common-model reassignment or by removing requests in decreasing order of $\sqrt{L_i}/d_i$, yielding $\boldsymbol a^0\in\A_t$.

\noindent\textbf{Step 2. Coordinated Potential Descent:}
Starting from $\boldsymbol a^0$, $\operatorname{Descend}$ alternates unilateral improvements and coordinated admission until no considered move decreases $F_t$ by more than $\eps$. The resulting profile $\boldsymbol a^L$ initializes the subsequent refinement.

\noindent\textbf{Step 3. Finite Binary Logit Refinement:}
Starting from $\boldsymbol a^L$, the algorithm performs $K_{\mathrm L}$ proposals according to~\eqref{eq:logit} and~\eqref{eq:temperature}. Feasible proposals are accepted probabilistically, while the best visited profile $\boldsymbol a^B$ and the final profile $\boldsymbol a$ are retained for further improvement.

\noindent\textbf{Step 4. Final Descent and Equilibrium Verification:}
The retained candidates undergo coordinated descent, and the incumbent is updated only when $F_t$ decreases by more than $\eps$. The returned assignment $\boldsymbol a^\star$ is then verified for feasibility, with its maximum feasible unilateral improvement measured by
\begin{equation}
	\mathcal E(\boldsymbol a)=
	\max_{\substack{i\in[N]\\
			\bar a_i\in\A_i(\boldsymbol a_{-i})}}
	\left[
	F_t(\boldsymbol a)
	-F_t(\bar a_i,\boldsymbol a_{-i})
	\right].
	\label{eq:gap}
\end{equation}
The unilateral gap characterizes the largest objective decrease attainable through a feasible unilateral deviation. The selected assignment is applied, and the model, checkpoint, and link-history states are updated for the next slot.

\subsection{Key Properties}\label{sec:properties}

We establish the principal properties of \SWAP, including admission priority, computational hardness, feasibility, and equilibrium guarantees. The following analysis considers a fixed decision slot with given request attributes, resource states, and service-window predictions.

\begin{proposition}[Admission priority]\label{prop:admission}
	For every feasible profile $\boldsymbol a\in\A_t$, the aggregate secondary loss satisfies $0\le G(\boldsymbol a)\le1$. For any two feasible profiles $\boldsymbol a,\boldsymbol b\in\A_t$, if $\boldsymbol b$ admits at least one more request than $\boldsymbol a$, then
$
	F_t(\boldsymbol b)-F_t(\boldsymbol a)
	\le-1+\eta<0.
	$
\end{proposition}
This property ensures that any feasible increase in admitted request cardinality takes precedence over variations in secondary performance costs.

\begin{proposition}[Computational hardness]\label{prop:hardness}
	The admission-priority scheduling problem in~\eqref{eq:objective} is NP-hard, even with two servers and one available inference model.
\end{proposition}

\begin{lemma}[Downward-closed feasibility]\label{lem:downward}
	For any feasible profile $\boldsymbol a\in\A_t$, replacing an arbitrary subset of admitted actions with rejection preserves feasibility under~\eqref{eq:cpu}. Moreover, the graph of feasible profiles connected by unilateral deviations is connected.
\end{lemma}
Removing admitted requests reduces computing contention without increasing model-storage occupancy or the service delays of retained requests. Thus, every feasible profile can be connected to the all-rejection profile, which also establishes the feasibility of the counterfactual in~\eqref{eq:utility}.

\begin{proposition}[Exact potential and finite descent]\label{prop:potential}
	For a fixed decision slot, $\Phi_t(\boldsymbol a)=-F_t(\boldsymbol a)$ is an exact potential function, and the constrained game admits at least one pure Nash equilibrium. Every sequence of strict feasible unilateral improvements terminates after finitely many updates. If $\mathcal E(\boldsymbol a)\le\eps$, the terminal profile constitutes a constrained pure $\eps$-Nash equilibrium.
\end{proposition}
The exact potential property follows from the marginal-contribution utility in~\eqref{eq:utility}, while finite termination follows from the finite feasible-profile set and strict objective descent.

\begin{proposition}[Finite refinement and incumbent preservation]\label{prop:refinement}
	Every coordinated descent stage in \SWAP\ terminates after finitely many accepted updates. Let $\boldsymbol a^L$ denote the initial descent solution and $\boldsymbol a^\star$ the returned profile. Then
	$
	\boldsymbol a^\star\in\A_t,\quad
	F_t(\boldsymbol a^\star)\le F_t(\boldsymbol a^L),\quad
	\mathcal E(\boldsymbol a^\star)\le\eps.
	$
\end{proposition}
Thus, the final assignment preserves feasibility and the approximate unilateral equilibrium condition without increasing the initial descent objective. It also admits no improvement greater than $\eps$ within the considered model-consolidation and two-request admission neighborhoods. Detailed proofs and computational complexity analysis are provided in Appendix~\ref{app:analysis}.

\section{Numerical Evaluations}\label{sec:experiments}

\subsection{Experimental Setting}\label{sec:settings}

We consider a UAV edge-inference network comprising $J=10$ ground-edge servers and $N\in\{20,40,60,80,100\}$ UAVs within a $1200\times1200$~m$^2$ area. 
UAVs follow random waypoint (RWP) mobility. The channel power gain is modeled as $g_{i,j}=g_0(d_{i,j}/d_0)^{-\nu}10^{X_{i,j}/10}$, where $d_{i,j}$ is the three-dimensional link distance, $g_0$ is the reference gain at distance $d_0$, $\nu$ is the path-loss exponent, and $X_{i,j}$ denotes zero-mean Gaussian shadowing in dB. The shadowing process follows first-order temporal correlation with coefficient $\exp(-0.2/10)$ between consecutive $0.2$~s samples. Each server provides heterogeneous computing capacity and shared model storage, and three inference models with different computation--quality profiles are considered. Each simulation episode consists of $100$ one-second slots, with the first $10$ slots used to initialize service states and causal histories. The remaining $90$ slots and $89$ consecutive transitions are used for performance evaluation. For each UAV load, all six methods are evaluated over the same $30$ independent environments, yielding $900$ method--environment episodes. 
The service-window predictor is trained using $82{,}959$ samples from $12$ environments, validated on $27{,}225$ samples from four environments, and calibrated using $27{,}217$ samples from another four environments. Each environment contains $20$ UAV trajectories. Three Quantile-GBDT regressors are implemented using LightGBM to estimate the remaining service-window quantiles. Training labels are obtained from subsequent channel observations over the capped prediction horizon, with trajectories extended beyond the episode boundary, while prediction features are derived from causal observations. The trained predictors and probability calibrators remain fixed throughout evaluation. All experiments are conducted on an Intel Core i9-12900H CPU, with scheduling decision times measured using one process and one numerical-library thread. Scheduling computation is excluded from the simulated physical execution timeline. The principal simulation and algorithm parameters are summarized in Table~\ref{tab:parameters}.

\begin{columntable}
	\caption{Simulation settings.}\label{tab:parameters}
	\centering
	\fontsize{8}{9.5}\selectfont
	\setlength{\tabcolsep}{2pt}
	\renewcommand{\arraystretch}{0.95}
	\begin{tabularx}{\columnwidth}{@{}>{\raggedright\arraybackslash}p{.30\columnwidth}|>{\raggedright\arraybackslash}X@{}}
		\hline
		\rowcolor[gray]{0.9}
		\textbf{Aspect} & \textbf{Setting}\\
		\hline
		Network Topology
		& $J=10$ servers; $N=20$--$100$ UAVs; area $1200\times1200$~m$^2$; altitude $80$~m.\\
		\hline
		Mobility and Channel
		& RWP; speed $5$--$15$~m/s; sampling interval $0.2$~s; $\nu=2.6$; $g_0=10^{-5}$ at $d_0=1$~m; shadowing SD $0.5$~dB; temporal correlation $\exp(-0.2/10)$.\\
		\hline
		Wireless Resources
		& Bandwidth $2$~MHz; power $0.2$~W; noise density $3.9811\times10^{-21}$~W/Hz; minimum rate $10$~Mbit/s; backhaul rate $100$~Mbit/s.\\
		\hline
		Edge Resources
		& CPU capacity Uniform $[8,14]$~GHz; model storage $100$~MB/server.\\
		\hline
		UAV Requests
		& Input size Uniform $[0.5,1.5]$~Mbit; computational intensity $[500,1000]$ cycles/bit; deadline $0.9$~s; checkpoint size Uniform $[0.2,1.0]$~MB; minimum quality $0.70/0.80$ with probabilities $0.65/0.35$.\\
		\hline
		Inference Models
		& Quality $(0.72,0.84,0.93)$; workload multiplier $(0.6,1.0,1.7)$; size $(20,40,70)$~MB; cold start $(0.08,0.16,0.28)$~s.\\
		\hline
		Service-Window Prediction
		& $T_{\max}=90$~s; quantiles $\{0.1,0.5,0.9\}$; LightGBM with $31$ leaves, $200$ trees, learning rate $0.06$, and minimum $50$ samples/leaf.\\
		\hline
		SWAP Configuration
		& $w_k=0.2$; $\eta=0.5$; $H=10$; $\delta=1$~s; $p=2$; $\eps=10^{-8}$; $K_{\mathrm L}=5N$; $\alpha=0.995$; $c_\tau=0.5$.\\
		\hline
		Evaluation Settings
		& $100$ slots/episode; $10$ initialization slots; $90$ evaluation slots; $30$ paired environments/load.\\
		\hline
	\end{tabularx}
\end{columntable}

\subsection{Benchmark Methods}\label{sec:baselines}

We compare \SWAP\ with five benchmark methods adapted from representative studies:
\textit{(i) WCOG}~\cite{wcog} adopts workload-ordered private-cost improvements for UAV computation offloading, with ground-edge execution and shared model selection.
\textit{(ii) SOTNRA}~\cite{sotnra} employs a server-pricing game with a finite price set and deterministic follower responses to coordinate computation offloading and resource allocation.
\textit{(iii) TWIST}~\cite{twist} integrates service caching and computation offloading through historical-demand-based cache updates, periodic model prefetching, and resource-aware assignment.
\textit{(iv) PG-EVC}~\cite{pgevc} performs random-order private-cost improvements based on current execution delay and inference-model quality.
\textit{(v) SGTO}~\cite{sgto} adopts a Stackelberg-based resource coordination mechanism with finite CPU-reservation choices and follower task-assignment optimization.
For fair comparison, all methods share the same UAV trajectories, wireless conditions, request attributes, server capacities, and model profiles, while maintaining their respective model-cache and session-checkpoint states. WCOG, SOTNRA, TWIST, and PG-EVC follow the computing allocation in~\eqref{eq:cpu}, whereas SGTO employs request-specific CPU reservations. The adapted objectives and scheduling rules are detailed in Appendix~\ref{app:baselinedetails}.

\subsection{Performance Metrics}\label{sec:metrics}

Performance metrics are first computed for each episode and then averaged across independent paired environments. We report $95\%$ Student's $t$ confidence intervals and assess paired differences using two-sided Wilcoxon signed-rank tests with Holm correction. The correction is applied separately by metric and load across five benchmark, nine ablation, or fourteen sensitivity comparisons, while prediction comparisons cover six mobility--load conditions per metric. We consider the following performance metrics.

\noindent\textit{1) Service Continuity and Admission:}
We use service continuity ratio (CSR), service-level agreement (SLA) satisfaction, and request drop ratio to evaluate continuous service provisioning. CSR measures successful transitions between consecutive evaluation slots, normalized by $89N$. A successful transition requires both requests to complete, the preceding wireless link to remain available until the next decision, and either an unchanged serving server or successful make-before-break migration and model initialization over the old connection. SLA satisfaction measures the fraction of requests completed within their deadlines and quality requirements over $90N$, while the drop ratio counts explicitly rejected requests over the same denominator. Wireless interruptions after admission affect CSR and SLA satisfaction but are distinguished from request rejection.

\noindent\textit{2) Service Efficiency and Resource Costs:}
We measure mean execution delay, the within-episode $95$th-percentile (P95) delay, and inference-model quality over completed requests. Model-deployment traffic is measured by the total size of newly loaded model instances per slot, including prefetched instances. The checkpoint migration rate is the number of inter-edge checkpoint transfers normalized by $89N$, including transfers following rejected slots. These metrics characterize execution performance and the resource costs associated with shared model deployment and service migration.

\noindent\textit{3) Prediction Accuracy and Computational Efficiency:}
Service-window prediction is evaluated using median-prediction mean absolute error (MAE), Brier score for calibrated interruption probabilities, and pinball losses at the $0.1$ and $0.9$ quantiles. Scheduling efficiency is assessed through per-slot decision time and the terminal unilateral gap $\mathcal E(\boldsymbol a^\star)$ defined in~\eqref{eq:gap}. Decision time measures the computational cost of scheduling, while the unilateral gap characterizes the approximate equilibrium attained by the returned assignment.

\subsection{Synthetic Experiments}\label{sec:synthetic}

\subsubsection{Impact of Request Load}\label{sec:mainresults}

\begin{figure*}[t]
	\centering
	\includegraphics[width=\textwidth]{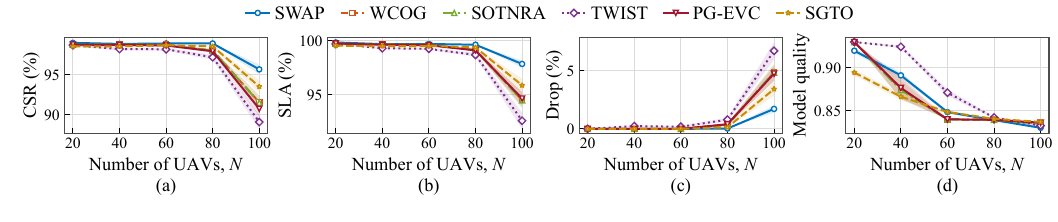}
	\caption{Service performance under varying UAV loads: (a) CSR, (b) SLA satisfaction, (c) request rejection, and (d) completed-request model quality. Curves and shaded regions represent episode means and 95\% confidence intervals over 30 paired environments.}
	\label{fig:service}
\end{figure*}

Figure~\ref{fig:service} illustrates the impact of increasing UAV loads on service performance. As shown in Fig.~\ref{fig:service}(a)--(c), all methods maintain relatively high CSR and SLA satisfaction with few rejected requests under light and moderate loads. As the number of UAVs increases further, resource contention becomes more pronounced, leading to declining service continuity and sharply increasing rejection rates. In contrast, \SWAP\ exhibits a smaller performance degradation and maintains the lowest rejection rate under heavy load. These results highlight the benefit of admission-priority coordination in adjusting server assignments and model selections to accommodate additional requests. The improvements in CSR, SLA satisfaction, and rejection are also statistically significant after Holm correction.

Figure~\ref{fig:service}(d) further shows that the average inference quality of completed requests decreases as the load increases, with the differences among methods gradually narrowing. This trend reflects the growing preference for computationally efficient model variants under resource contention. While \SWAP\ improves admission and service continuity, it does not consistently achieve the highest inference quality. Together, these observations illustrate how flexible model selection balances inference quality against the ability to serve more requests.

\subsubsection{Deployment and Migration Costs}

\begin{figure*}[t]
	\centering\vspace{-2mm}
	\includegraphics[width=\textwidth]{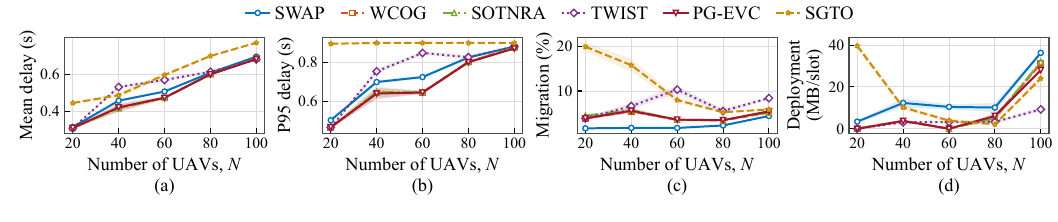}
	\caption{Execution delay and reconfiguration costs under varying UAV loads: (a) mean completed-request delay, (b) mean episode P95 delay, (c) checkpoint migration rate, and (d) model-deployment traffic. Shaded regions indicate 95\% confidence intervals over 30 paired environments.}
	\label{fig:resources}\vspace{-2mm}
\end{figure*}

Figure~\ref{fig:resources} examines execution delay and service reconfiguration costs under increasing request loads. As shown in Fig.~\ref{fig:resources}(a) and (b), both mean and P95 delays generally increase with the number of UAVs, reflecting intensified competition for shared computing resources. The differences among methods also vary with load, as their scheduling decisions affect both the admitted request population and the resulting computation delays.
Figure~\ref{fig:resources}(c) shows that \SWAP\ maintains comparatively low checkpoint migration rates across the evaluated loads, indicating that improved admission does not necessarily require frequent changes of serving servers. In contrast, model-deployment traffic exhibits a nonmonotonic trend, with a noticeable increase under the highest load, as illustrated in Fig.~\ref{fig:resources}(d). \SWAP\ incurs greater deployment traffic under congestion, whereas TWIST maintains lower model-loading traffic at the expense of higher request rejection. These results suggest that coordinated admission can exploit shared model deployment to serve additional requests while limiting session migration, although additional model instances may be required under severe resource contention.

\subsubsection{Service-Window Prediction Accuracy}\label{sec:predictionresults}

\begin{figure*}[t]
	\centering
	\includegraphics[width=\textwidth]{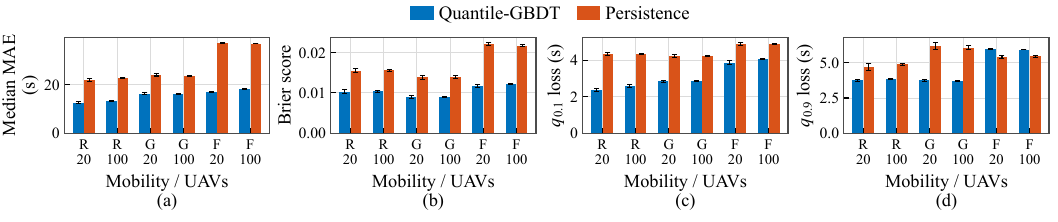}
	\caption{Service-window prediction performance: (a) median MAE, (b) Brier score, (c) 0.1-quantile pinball loss, and (d) 0.9-quantile pinball loss. Bars represent means and 95\% confidence intervals over 30 paired environments. R, G, and F denote RWP, Gauss--Markov, and faster RWP mobility, respectively.}
	\label{fig:prediction}
\end{figure*}

We compare Quantile-GBDT with a persistence predictor under RWP, Gauss--Markov, and faster RWP mobility, using identical accessible links and causal observations. The reference policy selects the nearest server satisfying $R_{i,j}\ge R_{\min}$, while the persistence predictor extrapolates recent transmission-rate trends and independently calibrates interruption probabilities.
As shown in Fig.~\ref{fig:prediction}(a) and (b), Quantile-GBDT consistently achieves lower median-prediction mean absolute error (MAE) and Brier scores across different mobility patterns and UAV loads. Both metrics generally deteriorate under faster mobility, reflecting the increased difficulty of predicting rapidly changing wireless connections. Nevertheless, Quantile-GBDT maintains its advantage over persistence, with statistically significant improvements after Holm correction.

The quantile results in Fig.~\ref{fig:prediction}(c) and (d) reveal a further distinction. Quantile-GBDT consistently reduces the lower-quantile pinball loss, supporting its use in estimating conservative service windows for shared model residence planning. However, its upper-quantile advantage diminishes and eventually reverses under faster mobility. These results indicate that service-window prediction provides more reliable central and lower-quantile estimates, while upper-tail predictions remain more sensitive to changes in mobility patterns.

\subsubsection{Ablation Study}\label{sec:ablation}

Table~\ref{tab:ablations} presents the ablation results for \SWAP\ and nine variants at $N=100$. Removing coordinated moves reduces CSR from 95.67\% to 93.86\%, a decrease of 1.81 percentage points, while increasing request rejection from 1.71\% to 2.76\%. The unilateral-descent variant exhibits a further reduction in CSR to 90.69\% and a rejection rate of 4.66\%. Restricting model selection to a fixed variant produces the lowest CSR of 90.03\% and the highest rejection rate of 4.89\%. These results demonstrate the importance of coordinated admission and flexible model selection under resource contention.
The contribution of finite binary logit refinement is also evident. Compared with the no-logit variant, \SWAP\ improves CSR from 95.33\% to 95.67\% and reduces rejection from 1.87\% to 1.71\%, with a corresponding increase in checkpoint migration from 3.91\% to 4.38\%. The CSR improvement is statistically significant at $N=100$ ($p_{\rm Holm}=1.83\times10^{-4}$), while no significant difference is observed at $N=60$ ($p_{\rm Holm}=1.000$).
The prediction-related variants further illustrate the role of service-window information. Replacing Quantile-GBDT with persistence slightly reduces CSR to 95.51\%, while removing the interruption-risk cost leads to a larger decrease to 94.89\%. Removing predictive planning, including the residence factor, further reduces CSR to 94.71\%. In contrast, replacing shared deployment charging with per-request charging has a comparatively small effect on continuity. The private-cost and unilateral-descent variants both exclude coordinated admission and logit refinement, and therefore reflect combined changes in the scheduling mechanism. Overall, the ablation results demonstrate the complementary roles of service-window prediction, coordinated admission, and finite logit refinement.
\begin{columntable}
	\caption{Ablation results at $N=100$.}
	\label{tab:ablations}
	\centering
	\fontsize{8.5}{10}\selectfont
	\setlength{\tabcolsep}{2pt}
	\renewcommand{\arraystretch}{1.08}
	\begin{tabularx}{\columnwidth}{@{}>{\raggedright\arraybackslash}Xrrrr@{}}
		\toprule
		Variant & CSR & Drop & Migr. & Deploy. \\
		\midrule
		SWAP & 95.67 & 1.71 & 4.38 & 36.30 \\
		Persistence predictor & 95.51 & 1.69 & 4.39 & 35.03 \\
		No risk cost & 94.89 & 1.72 & 4.37 & 36.48 \\
		No predictive planning & 94.71 & 1.76 & 4.83 & 31.14 \\
		Fixed model & 90.03 & 4.89 & 5.31 & 0.00 \\
		Unshared charge & 95.59 & 1.73 & 4.41 & 37.36 \\
		Private-cost descent & 90.80 & 4.66 & 2.88 & 27.78 \\
		Unilateral descent & 90.69 & 4.66 & 2.91 & 28.00 \\
		No logit refinement & 95.33 & 1.87 & 3.91 & 35.35 \\
		No coordinated moves & 93.86 & 2.76 & 5.67 & 37.52 \\
		\bottomrule
	\end{tabularx}
	\par\vspace{3pt}\raggedright
	\textit{Note: CSR, drop, and migration are percentages; deployment is measured in MB/slot. The private-cost and unilateral variants both exclude coordination and logit refinement.}
	\par
\end{columntable}

\subsubsection{Parameter Sensitivity}\label{sec:sensitivity}

We investigate the sensitivity of \SWAP\ to the interruption-risk weight, residence cap, quality exponent, logit temperature, proposal budget, and admission coefficient using ten paired environments per load. Increasing the interruption-risk weight generally improves service continuity, while varying the residence cap produces no statistically significant CSR difference after Holm correction. A higher logit temperature further improves continuity under heavy load, accompanied by increased checkpoint migration. At $N=100$, increasing $c_\tau$ from 0.5 to 8 raises CSR from 96.09\% to 96.41\% and migration from 4.33\% to 5.40\%, with both differences remaining significant after Holm correction ($p_{\rm Holm}=0.027$). The tested admission coefficients yield identical observed scheduling trajectories, indicating limited sensitivity within the evaluated range. These experiments use a separate set of environments from the main evaluation.

\subsubsection{Robustness to Mobility and Resource Changes}\label{sec:robustness}

\begin{figure*}[t]
	\centering
	\includegraphics[width=\textwidth]{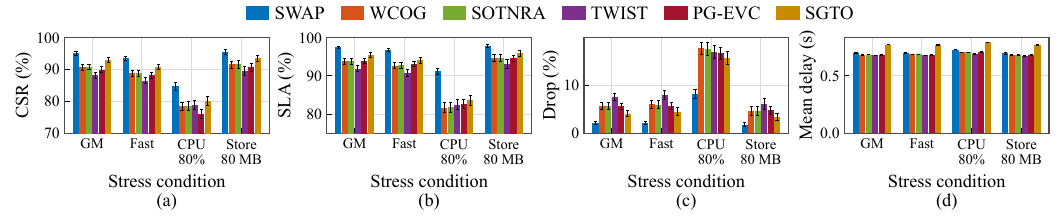}
	\caption{Robustness under mobility and resource variations at $N=100$: (a) CSR, (b) SLA satisfaction, (c) request rejection, and (d) completed-request delay. Gauss--Markov and faster RWP use 30 paired environments, while CPU and storage variations use ten. Error bars denote 95\% confidence intervals. The CSR and SLA axes begin at 70\% and 75\%, respectively.}
	\label{fig:robustness}
\end{figure*}

We further evaluate \SWAP\ under Gauss--Markov mobility, faster RWP motion, reduced CPU capacity (80\%), and reduced model storage (80~MB), retaining the original predictor and scheduling parameters. As shown in Fig.~\ref{fig:robustness}(a)--(c), \SWAP\ generally maintains higher service continuity and SLA satisfaction with fewer rejected requests across the tested conditions. Faster mobility reduces service continuity as wireless connections become less persistent, whereas reduced computing capacity causes a substantially larger deterioration in both service continuity and request admission.

Figure~\ref{fig:robustness}(d) further shows that completed-request delays remain relatively close across methods under different mobility patterns, while reduced CPU capacity introduces more pronounced delay increases. In comparison, reducing model storage has a smaller impact on service performance under the considered model configurations. These observations indicate that computing-resource scarcity is the dominant bottleneck in the evaluated scenarios, while predictive scheduling helps sustain service continuity under changing mobility conditions.

\subsection{Computational Efficiency and Discussion}\label{sec:computational}

All 15,000 \SWAP\ scheduling decisions in the main evaluation, including initialization slots, satisfy the feasible unilateral-gap condition $\mathcal E(\boldsymbol a^\star)\le\eps$, consistent with Proposition~\ref{prop:refinement}. Exhaustive enumeration of 40 small instances provides an additional check of the solution procedure. For 100 saved states at $N=100$, the median decision time is 0.666~s, with a P99 of 1.031~s and a maximum of 1.180~s. These results characterize the computational cost of coordinated scheduling, with scheduling time excluded from the simulated physical execution timeline.

Overall, \SWAP\ improves service continuity and request admission by integrating predicted link persistence, shared model deployment, and coordinated scheduling. The experimental results also highlight the associated tradeoffs in deployment traffic, checkpoint migration, inference quality, and execution delay. Future work will consider measured UAV trajectories and inference workloads, wireless resource contention, session expiration, and end-to-end scheduling overhead.

\section{Conclusion}\label{sec:conclusion}

This paper investigated joint request admission, edge-server assignment, and model selection for continuous UAV edge inference under uncertain wireless connectivity and shared resource constraints. We proposed Service-Window-Guided Admission and Model Placement (\SWAP), integrating service-window prediction, calibrated interruption risk, and residence-aware shared model deployment within an admission-priority framework. Based on marginal-contribution utilities, we constructed a constrained exact potential game and developed a scheduling algorithm combining unilateral improvement, coordinated admission, and finite binary logit refinement. Theoretical analysis established NP-hardness, equilibrium existence, and finite descent, while guaranteeing feasibility and an approximate unilateral equilibrium for the returned solution. Simulations against five adapted benchmarks demonstrated improved service continuity and reduced request rejection under resource contention, alongside tradeoffs in model-deployment traffic, checkpoint migration, inference quality, and execution delay. Prediction and ablation studies further highlighted the benefits of service-window awareness and coordinated scheduling, as well as sensitivity to faster mobility. Future work will consider measured UAV trajectories and inference workloads, wireless resource contention, scheduling overhead, and more flexible reassignment mechanisms.

\section*{Author Contributions}
\noindent Conceptualization: Houyi Qi and Minghui Liwang; methodology, software, experiments, and writing: Houyi Qi; review, editing, and supervision: Minghui Liwang. All authors have read and agreed to the published version of the manuscript.

\section*{Conflict of Interest}
\noindent All the authors declare that they have no conflict of interest.

	\begingroup\small
	\renewcommand{\bibfont}{\normalfont\fontsize{9.2}{11}\selectfont}
	\setlength{\bibsep}{4pt}
	\interlinepenalty=10000
	\bibliography{references}
	\endgroup
\end{multicols}
\clearpage
\begin{multicols}{2}
	\raggedcolumns
	\appendix

\section{Supplementary Analysis of SWAP}\label{app:analysis}

This appendix presents the computational complexity of \SWAP, proofs of its theoretical properties, and the formulations of the adapted benchmark methods.

\subsection{Computational Complexity}\label{app:complexity}

For a fixed decision slot, request attributes, inherited service states, and service-window predictions remain unchanged. A unilateral deviation affects only the source and destination servers, allowing their objective contributions to be updated independently of other servers. For fixed numbers of servers $J$ and models $M$, constructing or updating the counterfactual tables requires at most $O(N^2)$ operations, while scanning all unilateral candidates requires $O(N)$. A model-consolidation scan requires $O(N^2)$ operations, and a complete two-request admission scan requires at most $O(N^3)$. The finite binary logit stage has complexity $O(K_{\mathrm L}N^2)$, which reduces to $O(N^3)$ for $K_{\mathrm L}=O(N)$.

Let $U$, $C$, and $E$ denote the numbers of accepted unilateral updates, model-consolidation scans, and two-request admission scans across all descent stages, respectively. Excluding initialization, the per-slot complexity is bounded by $O(N^2(U+C+K_{\mathrm L}+1)+N^3(E+1))$, with $O(N^2)$ memory for counterfactual quantities. Under the default setting $K_{\mathrm L}=5N$, this simplifies to $O(N^2(U+C+1)+N^3(E+1))$.

During initialization, each server may examine up to $2^{M_j^{\rm inh}}$ subsets of inherited models, where $M_j^{\rm inh}\le M$ denotes the number of distinct models in its inherited assignment. Thus, initialization has exponential dependence on the model-library size when $M$ is unrestricted. Moreover, every accepted descent move decreases $F_t$ by more than $\eps$. Since $0\le F_t\le N+\eta$ and $\A_t$ is finite, the number of accepted updates within each descent stage is bounded by $\min\{|\A_t|-1,\lceil(N+\eta)/\eps\rceil\}$. Hence, every descent stage terminates after finitely many accepted updates.

\subsection{Proofs of Key Properties}\label{app:proofs}

\begin{proof}[Proof of Proposition~\ref{prop:admission}]
	For every admitted request, $0\le\widetilde D_i\le D_i\le d_i$, $0\le\pi_i\le1$, $\underline q_i\le q_i\le q_{\max}$, and $0\le\mu_iv_i\le v_{\max}$. Therefore, $G_D$, $G_R$, $G_Q$, and $G_M$ lie in $[0,1]$. Under~\eqref{eq:placement}, every resident server--model instance serves at least one admitted request, so the number of such instances is at most $N$. Since $H_{j,m}\ge1$ and $z_{j,m}s_m/H_{j,m}\le s_{\max}$, we also have $0\le G_P\le1$. Consequently, the aggregate secondary loss satisfies $0\le G(\boldsymbol a)\le1$ for all $\boldsymbol a\in\A_t$.
	
	If $\boldsymbol b$ admits at least one more request than $\boldsymbol a$, its rejection count decreases by at least one. Meanwhile, $G(\boldsymbol b)-G(\boldsymbol a)\le1$. It follows that $F_t(\boldsymbol b)-F_t(\boldsymbol a)\le-1+\eta<0$, establishing the strict admission-priority property.
\end{proof}

\begin{proof}[Proof of Proposition~\ref{prop:hardness}]
	We establish NP-hardness through a polynomial-time reduction from PARTITION~\cite{hardness}. Consider positive integers $\omega_1,\ldots,\omega_N$ with $\Omega=\sum_{i=1}^{N}\omega_i$. Construct an instance with two servers and one preloaded model, where $L_i=\omega_i^2$, $C_1=C_2=2\Omega^2$, $D_i^{\rm tx}=1/4$, and $d_i=1$. Choose $\Delta>1$ and feasible wireless-rate and model-quality parameters, with nonbinding storage constraints and zero migration and cold-start delays. Set $w_D=1$ and all other secondary weights to zero.
	
	For any assignment admitting all requests,~\eqref{eq:compute} yields $D_i^{\rm cmp}\le\omega_i/(2\Omega)\le1/2$. Hence, every such assignment satisfies the service deadline. By Proposition~\ref{prop:admission}, an optimal solution admits all $N$ requests.
	
	Let $\sigma$ denote the sum of the weights assigned to the first server, so the second server receives $\Omega-\sigma$. The aggregate computation delay is $[\sigma^2+(\Omega-\sigma)^2]/(2\Omega^2)$. Accordingly, the objective becomes
	\begin{equation}
		F_t=\frac{\eta}{N}
		\left[
		\frac{N}{4}
		+\frac{\sigma^2+(\Omega-\sigma)^2}{2\Omega^2}
		\right].
		\label{eq:partition}
	\end{equation}
	The objective attains the threshold $\frac{\eta}{N}(\frac{N}{4}+\frac14)$ if and only if $\sigma=\Omega/2$, which corresponds exactly to an equal-sum partition. Since the construction has polynomial encoding length, solving~\eqref{eq:objective} to optimality would solve PARTITION. Therefore, the proposed scheduling problem is NP-hard even with two servers and one model.
\end{proof}

\begin{proof}[Proof of Lemma~\ref{lem:downward}]
	Removing any subset of admitted requests decreases or preserves $\sum_{k\in\Sset_j}\sqrt{L_k}$ at each affected server. According to~\eqref{eq:compute}, the computation delay of every retained request cannot increase. Its transmission, migration, and cold-start delays remain unchanged, while model-storage occupancy cannot increase. Hence, the resulting assignment remains feasible.
	
	Every feasible profile can therefore reach the all-rejection profile through feasible unilateral removals. Conversely, reversing the removal sequence of any other feasible profile provides a feasible path from the all-rejection profile to that profile. Thus, the graph of feasible profiles connected by unilateral deviations is connected.
\end{proof}

\begin{proof}[Proof of Proposition~\ref{prop:potential}]
	Consider a feasible unilateral deviation from $\boldsymbol a$ to $\boldsymbol b=(\bar a_i,\boldsymbol a_{-i})$. Since the actions of all other players remain unchanged, the counterfactual term $F_t(0,\boldsymbol a_{-i})$ is identical for both profiles. Therefore,
	\begin{equation}
		\begin{aligned}
			u_i(\boldsymbol b)-u_i(\boldsymbol a)
			&=F_t(\boldsymbol a)-F_t(\boldsymbol b)=\Phi_t(\boldsymbol b)-\Phi_t(\boldsymbol a).
		\end{aligned}
		\label{eq:potentialidentity}
	\end{equation}
	Hence, $\Phi_t=-F_t$ is an exact potential function. Since $\A_t$ is finite and nonempty, it contains a minimizer of $F_t$, which constitutes a constrained pure Nash equilibrium. Every strict feasible unilateral improvement decreases $F_t$, so the same profile cannot be revisited and the descent process terminates after finitely many updates. If no feasible unilateral deviation yields an improvement greater than $\eps$, the terminal profile satisfies $\mathcal E(\boldsymbol a)\le\eps$ and is therefore a constrained pure $\eps$-Nash equilibrium.
\end{proof}

\begin{proof}[Proof of Proposition~\ref{prop:refinement}]
	Every accepted unilateral or coordinated update preserves feasibility and reduces $F_t$ by more than $\eps$. Since $\A_t$ is finite, each coordinated descent stage terminates after finitely many accepted updates. Its final exhaustive unilateral scan establishes $\mathcal E(\boldsymbol a)\le\eps$, including for the initial descent solution $\boldsymbol a^L$.
	
	The subsequent binary logit stage performs $K_{\mathrm L}$ proposals, after which the retained candidates undergo coordinated descent. The incumbent is initialized at $\boldsymbol a^L$ and replaced only when a refined candidate improves its objective by more than $\eps$. Consequently, the returned profile satisfies $\boldsymbol a^\star\in\A_t$, $F_t(\boldsymbol a^\star)\le F_t(\boldsymbol a^L)$, and $\mathcal E(\boldsymbol a^\star)\le\eps$.
\end{proof}

\subsection{Adapted Benchmark Formulations}\label{app:baselinedetails}

The benchmark methods retain their respective computation-offloading, service-caching, and resource-allocation principles under the common service constraints. Let $A_i=b_i\phi_i$ denote the nominal computation workload and $\chi_{i,j}$ a dimensionless server-dependent fee. The following formulations specify the adapted objectives and scheduling rules used in the experiments.

\noindent\textit{WCOG and PG-EVC:}
For an admitted request assigned to server $e_j$, both methods use the current-state private cost
\begin{equation}
	c_i^{\rm cur}(\boldsymbol a)
	=\frac{D_i(\boldsymbol a)}{d_i}
	+\frac{q_{\max}-q_i}{q_{\max}-\underline q_i}
	+\chi_{i,j}.
	\label{eq:baselinecost}
\end{equation}
The first two terms represent normalized service delay and inference-quality loss. Rejection incurs $c_i^{\rm cur}(0,\boldsymbol a_{-i})=\rho_{\rm rej}$, where $\rho_{\rm rej}=8N$, and both methods set $\chi_{i,j}=0$. For each candidate assignment, the associated delay, workload, and model quality are updated, and the selected request adopts its minimum-cost feasible action. WCOG prioritizes improving requests with larger nominal workloads $A_i$, whereas PG-EVC selects uniformly among improving requests. Each accepted update decreases the private cost by more than $\eps$. The process terminates when no improving action remains or after at most $10{,}000$ accepted updates.

\noindent\textit{SOTNRA:}
The follower adopts~\eqref{eq:baselinecost} with the server-dependent fee $\chi_{i,j}=p_jA_i/(C_j\Delta)$, where $p_j$ is selected from the finite price set $\mathcal P_{\mathrm S}$. The leader's utility is
\begin{equation}
	V_j=\sum_{i\in\Sset_j}\chi_{i,j}
	-\lambda_{\mathrm S}
	\left(
	\frac{\sum_{i\in\Sset_j}L_i}{C_j\Delta}
	\right)^2.
	\label{eq:pricegame}
\end{equation}
Here, $\lambda_{\mathrm S}>0$ weights the normalized computing-load cost. We set $\mathcal P_{\mathrm S}=\{0,0.1,0.2,0.4,0.8\}$ and $\lambda_{\mathrm S}=0.1$. Each price vector induces a deterministic follower response from the same unpriced, workload-ordered initial assignment. Servers update their prices sequentially and accept improvements greater than $\eps$. The process terminates upon an unchanged sweep, a repeated price vector, or the completion of $20$ sweeps, returning the follower assignment under the final prices.

\noindent\textit{TWIST:}
The caching mechanism maintains discounted model demand $D_{j,m}^{\rm hist}$ and virtual offered-load pressure $Q_j^{\rm v}$, both initialized at zero. Hypothetical assignments satisfying the uplink-rate and model-quality requirements estimate offered demand without enforcing service deadlines. These assignments use~\eqref{eq:baselinecost} with $\chi_{i,j}=\lambda_{\mathrm T}^{\rm fee}Q_j^{\rm v}A_i/(C_j\Delta)$.

Let $o_{j,m}$ denote the number of hypothetical assignments to server--model pair $(j,m)$ and $O_j$ their total model-dependent workload at server $j$. The historical states are updated as $D_{j,m}^{\rm hist,+}=\theta_{\mathrm T}D_{j,m}^{\rm hist}+o_{j,m}$ and $Q_j^{\rm v,+}=\max\{0,Q_j^{\rm v}+O_j/(C_j\Delta)-1\}$, where $\theta_{\mathrm T}\in(0,1)$ is the demand-discount factor.

Every ten slots, the cache planner selects a storage-feasible model-residency plan $Y=\{Y_{j,m}\}$ by minimizing
\begin{equation}
	\begin{aligned}
		\Psi(Y)=&
		\sum_{i=1}^{N}c_i^{\rm cur}(\boldsymbol a(Y))+\lambda_{\mathrm T}^{\rm load}
		\frac{\sum_{j,m}Y_{j,m}(1-y^-_{j,m})s_m}
		{C_{\rm norm}}\\
		&-\lambda_{\mathrm T}^{\rm hist}
		\frac{\sum_{j,m}D_{j,m}^{\rm hist}Y_{j,m}}
		{\max\{1,\sum_{j,m}D_{j,m}^{\rm hist}\}}.
	\end{aligned}
	\label{eq:cachescore}
\end{equation}
Here, $\sum_{j,m}$ denotes summation over $[J]\times[M]$, $Y_{j,m}\in\{0,1\}$ indicates planned model residency, and $\boldsymbol a(Y)$ is the follower assignment restricted to the cached models. The normalization constant is $C_{\rm norm}=\max\{Ns_{\max},\sum_jS_j\}$. The coefficients $\lambda_{\mathrm T}^{\rm fee}$, $\lambda_{\mathrm T}^{\rm load}$, and $\lambda_{\mathrm T}^{\rm hist}$ weight the virtual-load fee, model-loading cost, and historical-demand preference, respectively.

We set $\theta_{\mathrm T}=0.9$ and $\lambda_{\mathrm T}^{\rm fee}=\lambda_{\mathrm T}^{\rm load}=\lambda_{\mathrm T}^{\rm hist}=0.1$. Each cache refresh performs at most three improvement sweeps. The selected cache plan remains fixed between refreshes, while newly loaded model instances contribute to deployment traffic and incur their full cold-start delays.

\noindent\textit{SGTO:}
Each server $e_j$ reserves computing rates $\bar f_{i,j}\ge0$ satisfying $\sum_{i=1}^{N}\bar f_{i,j}\le C_j$. For a positive reservation, the computation delay is $L_{i,m}/\bar f_{i,j}$. Define $\widehat f_{i,j}=\bar f_{i,j}/10^9$ and $\widehat L_{i,m}=L_{i,m}/10^9$ as the numerical computing rate in GHz and workload in Gcycles, respectively. The normalized computing price is $p_{i,j}^{\rm cpu}=\kappa_{\rm p}\widehat f_{i,j}/\widehat f_{\rm ref}$, where $\kappa_{\rm p}>0$ is the pricing coefficient and $\widehat f_{\rm ref}>0$ is the reference computing rate in GHz.

For a candidate assignment $(j,m)$, let $D_i(j,m)$ denote the full service delay under the fixed reservations. The follower cost is
\begin{equation}
	\begin{aligned}
		c_i^{\rm SGTO}(j,m)=
		\frac{D_i(j,m)}{d_i}
		+\frac{q_{\max}-q_m}{q_{\max}-\underline q_i}+p_{i,j}^{\rm cpu}\widehat L_{i,m}
		+\lambda_{\rm tx}\frac{D_i^{\rm tx}}{D_{\rm ref}}.
	\end{aligned}
	\label{eq:sgtocost}
\end{equation}
Here, $\lambda_{\rm tx}\ge0$ weights normalized transmission delay and $D_{\rm ref}>0$ is the reference duration. Rejection incurs $\rho_{\rm rej}$, while the leader maximizes $V_j^{\rm SGTO}=\sum_{i\in\Sset_j}(p_{i,j}^{\rm cpu}-\kappa_{\rm c})\widehat L_i$, where $\widehat L_i=L_i/10^9$ and $\kappa_{\rm c}\ge0$ is the computing-cost coefficient. We set $\kappa_{\rm p}=0.03$, $\widehat f_{\rm ref}=0.2$, $\lambda_{\rm tx}=0.25$, $D_{\rm ref}=1$~s, and $\kappa_{\rm c}=0.002$.
Each server considers at most nine computing-reservation vectors based on its initial allocation and alternative distributions of available capacity. Let superscript $0$ denote quantities evaluated at SGTO's feasible initial assignment. For an initially admitted request, $L_i^0$ is its workload, $f_i^0>0$ is its computing share, and $D_i^{\rm nc,0}=D_i^{\rm tx,0}+D_i^{\rm mig,0}+D_i^{\rm cold,0}$ is its initial noncomputing delay. For a request initially assigned to server $e_j$, the reservation floor is
\begin{equation}
	\bar f_{i,j}^{\min}
	=\min\!\left\{
	f_i^0,\frac{L_i^0}{d_i-D_i^{\rm nc,0}}
	\right\}.
	\label{eq:sgto-reservation-floor}
\end{equation}
Initial feasibility ensures $d_i-D_i^{\rm nc,0}\ge L_i^0/f_i^0>0$. The reservation floor is zero for requests not initially assigned to server $e_j$.
Under fixed reservations, the follower minimizes the sum of admitted-action costs and rejection penalties over binary assignments, subject to the rate, quality, delay, and storage constraints. The selected assignments determine model occupancy through~\eqref{eq:placement}, while unused computing reservations are not redistributed. Server-ordered leader updates terminate upon cycle detection or after at most $20$ sweeps.

\end{multicols}
\end{document}